# Strong suppression of near-field radiative heat transfer by superconductivity in NbN

Authors: Věra Musilová, Tomáš Králík, Tomáš Fořt and Michal Macek

*The Czech Academy of Sciences, Institute of Scientific Instruments, Královopolská 147, 612 64 Brno, Czech Republic*

Abstract:

Near-field (NF) radiative heat transfer (RHT) over vacuum space between closely spaced bodies can exceed the Planck's far-field (FF) values by orders of magnitude. A strong effect of superconductivity on NF RHT between plane-parallel thin-film surfaces of niobium (Nb) was recently discovered and discussed in a short paper [1]. We present here an extensive set of experimental results on NF as well as FF RHT for geometrically identical samples made of niobium nitride (NbN), including a detailed discussion of the experimental setup and errors. The results with NbN show more precise agreement with theory than the original experiments with Nb.

We observed a steep decrease of the heat flux at the transition to superconductivity when the colder sample (absorber) passed from the normal to the superconducting (SC) state ($T_c$≈15.2 K), corresponding to an up to eightfold contrast between the normal and SC states. This differs dramatically from the situation in the FF regime, where only a weak effect of superconductivity was observed. Surprisingly, the contrast remains sizable even at high temperatures of the hot sample (radiator) with the characteristic energy of radiation far above the SC energy gap. We explain the maximum of contrast in heat flux between the normal and SC states, found at a distance about ten times shorter than the crossover distance between NF and FF heat flux, being $d$≈1000/$T$ [µm]. We analyze in detail the roles of transversal electric (TE) and magnetic (TM) modes in the steep decrease of heat flux below the SC critical temperature and the subsequent flux saturation at low temperatures.

Interestingly, we expose experimentally the effect of destructive interference of FF thermal radiation in the vacuum gap, which was observable at temperatures below the absorber superconducting transition.

## I. INTRODUCTION

The electromagnetic near field (NF), in contrast to the far field (FF), is not radiative. However, when another body is placed sufficiently near to the surface of a thermal source, it can absorb a significant amount of power of the thermal evanescent waves by photon tunneling [2]. This NF regime of heat transfer potentially exceeds the Planck's blackbody limit by orders of magnitude [3], [4], [5]. It occurs prevalently by photons of much lower energy than the FF. A crossover between FF and NF regimes occurs when the distance $d$ from the source reaches ~$10^{-1}$ of the radiation wavelength $\lambda$ [6]. The NF heat transfer has been studied for dielectrics [7], [8], [9] metals [4], [10], [5], materials undergoing metal-to-insulator transition [11], metamaterials [12,13], and graphene [14], [15], [16], to give some examples. As outlined in [1], the effect may become of practical significance in thermo-voltaics, thermal NF microscopy [17], or in contactless thermal control in microelectromechanical devices [18]. For a review see [2], [12] and [19]. Most of the NF studies have been theoretical due to substantial challenges in experiments [4,8,11,20-22], which originate in very small distances necessary to reach the NF heat transfer regime between samples. Relevant scales for the vacuum gap at room (cryogenic) temperatures are nm (µm), thus any thermal dilatation of the samples causes a considerable parasitic heat conduction risk.

In a recent short paper [1], we inspected the effect of the metal-to-superconductor (SC) transition on both NF and FF heat transfer. We found experimentally that when the colder sample (absorber) transits to the SC state, the power transferred by the NF component strongly decreases, while the effect on the FF radiation component is very weak. This holds even if the hotter sample (radiator) temperature $T_2$ is much higher than $T_c$.

Agreement between theory and the data for normal and superconducting Nb planar surfaces with a parallel vacuum gap between them was only qualitative in Ref. [1], apparently due to the samples' imperfection, caused by the high reactivity of Nb. Therefore, we performed detailed experiments in the same setup, but dedicatedly on a different material–the niobium nitride NbN–which is reported to obey the BCS theory well in the SC state [23]. The results are presented here, in a more extensive form and together with experimental and numerical details omitted in [1].

We could identify in both the experiment and theory, that the transverse electric (TE) mode (electrical field parallel to the surfaces) is dominating the NF heat exchange between the plane-parallel surfaces of metals in the normal state. Crucially, as distinct from the transverse magnetic (TM) mode, the TE component is substantially suppressed in the SC state (possibly surrendering to the TM mode contribution), causing thus the pronounced decrease in the NF heat transfer, which represents our key result. Suppression of the NF TE mode by superconductivity also uncovers the region of destructive interference of thermal radiation in the FF heat transfer regime. This manifests itself by a minimum of heat transfer at a specific distance, close to the NF-FF crossover, which was unobservable in the normal state. With the data obtained for NbN, contrary to the Nb [1], we can experimentally prove more decisively even the above detailed effects of the SC transition on the near field radiative heat transfer.

The paper is organized as follows: Section II presents theoretical relations for radiative heat transfer between plane parallel surfaces and the model describing the samples used in experiments. Section III contains the experimental details, including the measurement uncertainty. The results in Section IV contain representative examples of the measured heat power transferred between the samples depending on their temperatures (crossing the superconducting transition) and on their distance. In Sec. V, we analyze and compare the measured results with theory: In particular, we discuss the effect of superconducting transition on NF heat flux (Sec. V. A), interpret the heat conductivities of the vacuum gap between the samples (Sec. V. B) and the distance and temperature dependences of the NF heat transfer in terms of emissivity (Sec. V. C), show the roles of TE and TM modes in NF and FF radiative heat transfer above and below the superconducting transition (Sec. V. D), and discuss the heat radiation interference (Sec. V. E). Section VI summarizes the results and brings conclusions.

In the Supplemental Material [24], we present calculations of the optical constants and the transmissivities relevant for our experiment and analyze in detail the influences of parameters in the theoretical model of the samples.

## II. THEORY

### A. Radiative heat transfer equations

For layered samples in plane parallel configuration used in this experimental work we can calculate the NF heat transfer directly from the relations derived by Polder and Van Hove [3] for infinite plane-parallel, homogeneous and isotropic surfaces. We briefly review the relevant relations below:

The total heat flux over the vacuum gap from the radiator to the absorber reads

$$q(T_1,T_2,d) = \int_0^\infty d\omega\, I(\omega,T_1,T_2)\,\mathcal{T}(\omega,d,T_1,T_2) \qquad (1)$$

which contains the intensities $I$ of the blackbody radiation generated within the bodies at temperatures $T_1$ and $T_2$

$$I(\omega,T_1,T_2) = \frac{1}{\pi}\left(\frac{\omega}{2\pi c}\right)^2 \left[\frac{\hbar\omega}{\exp(\hbar\omega/k_B T_2)-1} - \frac{\hbar\omega}{\exp(\hbar\omega/k_B T_1)-1}\right] \qquad (2)$$

and the spectral hemispherical transmissivity $\mathcal{T}$ of the vacuum gap. The latter can be decomposed as $\mathcal{T}(\omega,d,T_1,T_2) \equiv \mathcal{T}_{TE}^{FF} + \mathcal{T}_{TM}^{FF} + \mathcal{T}_{TE}^{NF} + \mathcal{T}_{TM}^{NF}$ into the transmissivities for transverse electric mode (TE) and transverse magnetic (TM) modes for both the FF and NF components, given by

$$\mathcal{T}_{TE,TM}^{FF}(\omega,d,T_1,T_2) = \frac{2\pi}{(\omega/c)^2}\int_0^{\omega/c} \frac{1}{2}\, t_{TE,TM}^{FF} K\, dK \qquad (3a)$$

$$\mathcal{T}_{TE,TM}^{NF}(\omega,d,T_1,T_2) = \frac{2\pi}{(\omega/c)^2}\int_{\omega/c}^{\infty} \frac{1}{2}\, t_{TE,TM}^{NF} K\, dK \qquad (3b)$$

where $K$ is the surface-parallel component of the wave vector $\boldsymbol{K_0}$ in vacuum of magnitude $|\boldsymbol{K}_0|=\omega/c \equiv K_0$, and $t_{TE,TM}^{NF,FF}$ are the spectral directional NF and FF transmissivities for the TE and TM modes of the vacuum gap between plane-parallel samples. These are given in terms of the Fresnel reflection coefficients $r$ of the samples (multilayers in general),

$$t_m^{FF} = \frac{\left(1-|r_m^{(1)}|^2\right)\left(1-|r_m^{(2)}|^2\right)}{\left|1-r_m^{(1)}\, r_m^{(2)}\exp(2\,i\gamma_0 d)\right|^2}, \qquad (4a)$$

$$t_m^{NF} = \frac{4\,\mathrm{Im}(r_m^{(1)})\,\mathrm{Im}(r_m^{(2)})\exp(-2\gamma_0'' d)}{\left|1-r_m^{(1)} r_m^{(2)}\exp(-2\gamma_0'' d)\right|^2}, \quad m = TE, TM \qquad (4b)$$

and are in Eq. (3) integrated over the angles. The superscript distinguishes the absorber (1) and radiator (2) samples. The exponential terms in Eq. (4) are of "interference form" for FF, and of decaying form for NF, with the real values $\gamma_0=[(\omega/c)^2 - K^2]^{1/2}$ ($K<K_0$) and $\gamma_0''=[K^2 - (\omega/c)^2]^{1/2}$ ($K>K_0$) multiplied by the intersample distance $d$.

As an alternative, we rewrite Eq. (1) as

$$q(T_1,T_2,d) = \int M(\omega,d,T_1,T_2)\,[E_2(\omega,T_2) - E_1(\omega,T_1)]\,d\omega, \qquad (5)$$

$$E_i(\omega,T_i) \equiv \frac{\hbar\omega}{\exp(\hbar\omega/k_B T_i)-1}, \quad i=1,2, \qquad (6)$$

$$M(\omega,d,T_1,T_2) \equiv \frac{1}{\pi}\left(\frac{\omega}{2\pi c}\right)^2 \mathcal{T}(\omega,d,T_1,T_2) \qquad (7)$$

in which $E$ represents the energy of a harmonic oscillator at temperature $T_i$ and $M$ is the „modified transmissivity", decomposed similar to $\mathcal{T}$ as $M(\omega,d,T_1,T_2) \equiv M_{TE}^{FF} + M_{TM}^{FF} + M_{TE}^{NF} + M_{TM}^{NF}$. The energy $E$ is strictly monotonic in both $\omega$ as well as $T$ and thus $M$ preserves the essential features of the heat flux spectrum. Such separation will be useful in the discussion of results in Sec. V D.

### B. Near field heat transfer and superconductivity

An intuitive physical explanation of the effect discussed in this paper is as follows: The optical constants of a BCS superconductor, and consequently the transmissivities $M$, or $\mathcal{T}$, vary strongly with temperature at frequencies within the energy gap. Taking into account the BCS frequency value [25] for a superconductor,

$$\hbar\omega_{g0} = 3.528\, kT_c \qquad (8)$$

and the "median frequency" $\omega_m$ of the Planck's energy spectrum at radiator temperature $T_2$

$$\hbar\omega_m \approx 3.5\, kT_2 \quad (\lambda_m = 2\pi c/\omega_m \approx 4100/T_2\ [\mu\text{m;K}]), \qquad (9)$$

we infer from the relation $\omega_m \approx \omega_{g0}$ that for a radiator at $T_2>T_c$, less than half of the radiated energy falls within the energy gap (supposing a weak frequency dependence of the radiator emissivity). Then the transition of the absorber to superconductivity ($T_1<T_c$) will affect the FF heat transfer between the metallic bodies by a factor of less than 2.

On the other hand, at intersample distances $d<1000/T_2 \approx \lambda_m/4$ [µm; K], the NF heat transfer dominates the FF [1,5]. As the NF heat transfer occurs predominantly at wavelengths longer than $d$, the dominating wavelengths of the NF may fall into the energy gap even at radiator temperatures $T_2>T_c$. As a result, the transition of the absorber to BCS superconductivity should strongly suppress the heat transfer in the NF regime.

### C. Model of the samples

To calculate optical properties of the NbN samples, we applied the Drude model, $\sigma = \sigma_{DC} / (1 - i\omega\tau)$, $\tau = \sigma_{DC} / (\varepsilon_0 \omega_p^2)$ and the published value [26] of the plasma frequency $\omega_p = 14.7 \times 10^{15}\,\mathrm{rad/s}$, together with the Mattis-Bardeen weak-coupling BCS theory [27], [28] of the dynamic electrical conductivity $\sigma(\omega)$. Due to short electronic relaxation time $\tau$ of NbN and correspondingly high value of impurity parameter $y = \hbar / 2\Delta\tau$ in the BCS model [28], dc conductivity $\sigma_{dc}$ and temperature $T_c$ of superconducting transition are decisive in the description of both normal and superconducting NbN in our low-temperature heat transfer experiment (Supplemental Material [24]). The measured resistivity $\rho$ related to $\sigma_{dc}$ is depicted in Fig. 1 together with the critical temperature $T_c$.

For the SC energy gap at absolute zero, we applied the relation

$$\hbar\omega_{g0} = 2nkT_c\,, \quad n = 1.95. \tag{10}$$

The value $n$=1.95 is supported by published experimental results on the energy gap of NbN films with thickness and critical temperature similar to our samples (50 nm, $T_c$=15.8 K [29] and 250nm, $T_c$=13.5 K [30]). The temperature dependence of the energy gap $E_{gT}$ is taken in accord with the BCS theory. Its values at temperatures $0<T<T_c$ were obtained by interpolation in the table of reduced energy gaps $E_{gT}/E_{g0}$ vs $T/T_c$ taken from [31].

We performed the calculations for a multilayer system of samples consisting of a $L$=270 nm thick NbN layer on a 2.7 mm thick sapphire plate covered with a nontranslucent metallic layer on its reverse side (cf. Sec. III. A). We approximate the sapphire layer with the optical constants of the ordinary ray at room temperature [32]. As discussed in detail in the Appendix, we tested the effects of the sapphire substrate and the NbN layer thickness $L$ by comparing heat fluxes computed for the multilayer model with results for the respective NbN layers placed in vacuum without any substrate. The differences we found were only within a few percent. We conclude that the sapphire substrate influences the theoretical heat flux transferred over the vacuum gap between the samples here studied only very weakly. We note that the thickness of the NbN samples presented here is nearly optimal (see the Supplemental Material [24]) for reaching the maximum effect of the superconducting transition on the NF heat transfer. For sensitivity of the theoretical model of the NF heat flux to the sample parameters, see the Supplemental Material [24].

## III. EXPERIMENTAL SETTING

### A. Samples preparation and properties

Both the radiator and absorber samples are $L$=270 nm thick NbN layers deposited by dc magnetron sputtering on sapphire circular substrates 35 mm wide and 2.7 mm thick, with planarity better than 0.5 µm, and appearing optically smooth. The sputtering procedure was optimized to reach as high a $T_c$ as possible by tuning the concentration of nitrogen in the Ar atmosphere in the deposition chamber. The reverse side and sidewalls of the sample substrates are covered with auxiliary ~1µm-thick layers of Al and Cu, as in the experiments of Ref. [1]. The Al layer suppresses the absorption or emission of parasitic background radiation. The Cu "patches" serve as contacting electrodes for in situ capacitive measurements of the vacuum gap distance $d$ between the NbN sample surfaces, which are always in concentric plane-parallel position.

Figure 1 shows the dc resistivity of the samples, as measured by a four-point probe. The measurements resulted in $\rho$-values slightly increasing towards low temperatures, $\rho$(300 K)/$\rho$(16 K)=0.93, and achieving nearly constant values of residual resistivity, both at temperatures ranging from 50 K down to 16 K. Upon cooling both of the samples further, we observe transitions to the SC state at the following $T_c$ values:

$$T_{c1} = (15.25 \pm 0.01)\ \mathrm{K} \quad \text{(absorber)}, \tag{11a}$$

$$T_{c2} = (15.21 \pm 0.01)\ \mathrm{K} \quad \text{(radiator)}, \tag{11b}$$

which mutually agree within an interval of ~0.1 K. The values of $T_{c1}$ and $T_{c2}$ are determined at temperature points corresponding to half the value of residual resistivity $1.170\times10^{-6}\ \Omega\mathrm{m}$ (absorber) and $1.177\times10^{-6}\ \Omega\mathrm{m}$ (radiator). The four-point probe measurements were conducted just before installation of the samples into the NF heat transfer apparatus EWA (evanescent wave apparatus), described in the next section.

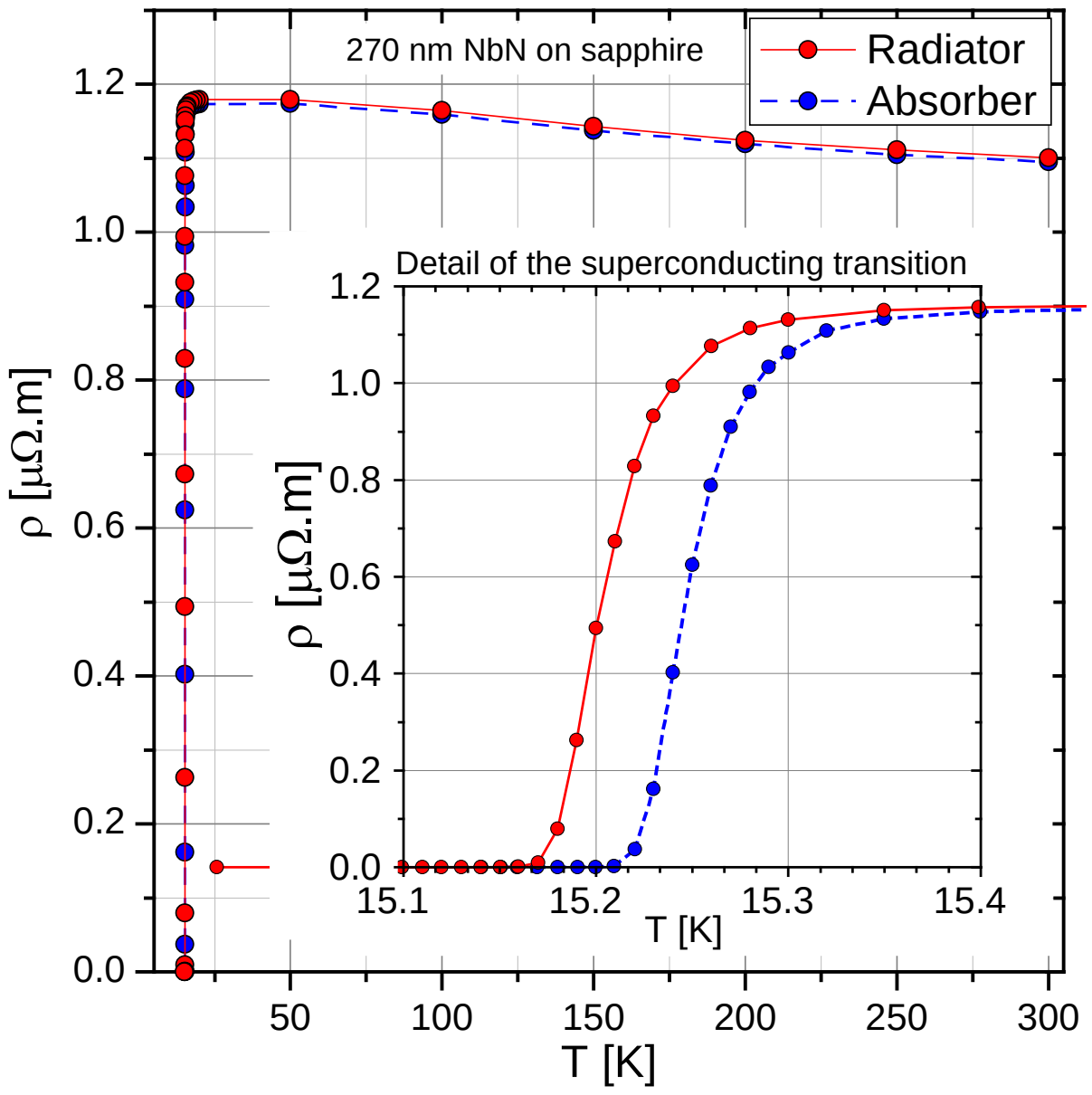

Fig. 1. (Color online) Resistivity dependences on the temperatures of the absorber and radiator NbN samples. We determine the critical temperatures at half the maximum value of residual resistivity: for the absorber $T_{c1}$=(15.25±0.01) K, and for the radiator $T_{c2}$=(15.21±0.01) K. Decrease from the 95 % down to 5 % of residual resistivity value occurs within a temperature interval of about 120 mK, see the inset. We measured the resistivity by a four-point probe at the samples' center.

Residual resistivity provide the values of the samples' dc conductivity used in the BCS calculation as

$$\sigma_{DC1} = 8.50\times10^5 \text{ S}, \quad (12a)$$

$$\sigma_{DC2} = 8.55\times10^5 \text{ S}. \quad (12b)$$

### B. Heat transfer measurement

Figure 2 shows the core part of the "evanescent wave apparatus" (EWA) used for the NF heat transfer measurements. For experiments with SC samples, EWA has been updated by an additional heater for setting stabilized absorber temperatures $T_1$ in the range from 5 to 20 K; see also [1]. This enables us in particular to vary and cross $T_c$ with the absorber temperature $T_1$, in addition to the radiator temperature $T_2$, variable already in the EWA version of Refs. [5,33].

We measure the heat power $P$ transferred from the radiator, stabilized at temperature $T_2$ with a heat flow meter (HFM, thermal resistor), which makes a connection between the absorber and a support kept at a stabilized temperature $T_0$. The temperature $T_1$ of the absorber equilibrates at a value slightly higher than $T_0$. We infer the value of $P$ from the temperature drop $T_1$-$T_0$ on the HFM. As the HFM is in situ calibrated using an electrical heater, the accuracy of the measured heat flow $P$ is determined only by the resolution of temperatures $T_0$ and $T_1$ (not by the accuracy of the thermometers used), the thermometers reproducibility, the uncertainty in the calibrating electrical power, and the electrical stability. Resolutions of the temperature measurements, which are $T_{0,\,res}$=50 μK, $T_{1,\,res}$=50 μK (for $T_1$<10 K) and 500 μK (for $T_1$>10 K) give ~ 0.1 μW as the lowest value of P with an uncertainty of a few percent, read out from the calibration curve $P_{cal} \approx 3.8 T_0^{1.222}(T_1 - T_0)$ [μW; K].

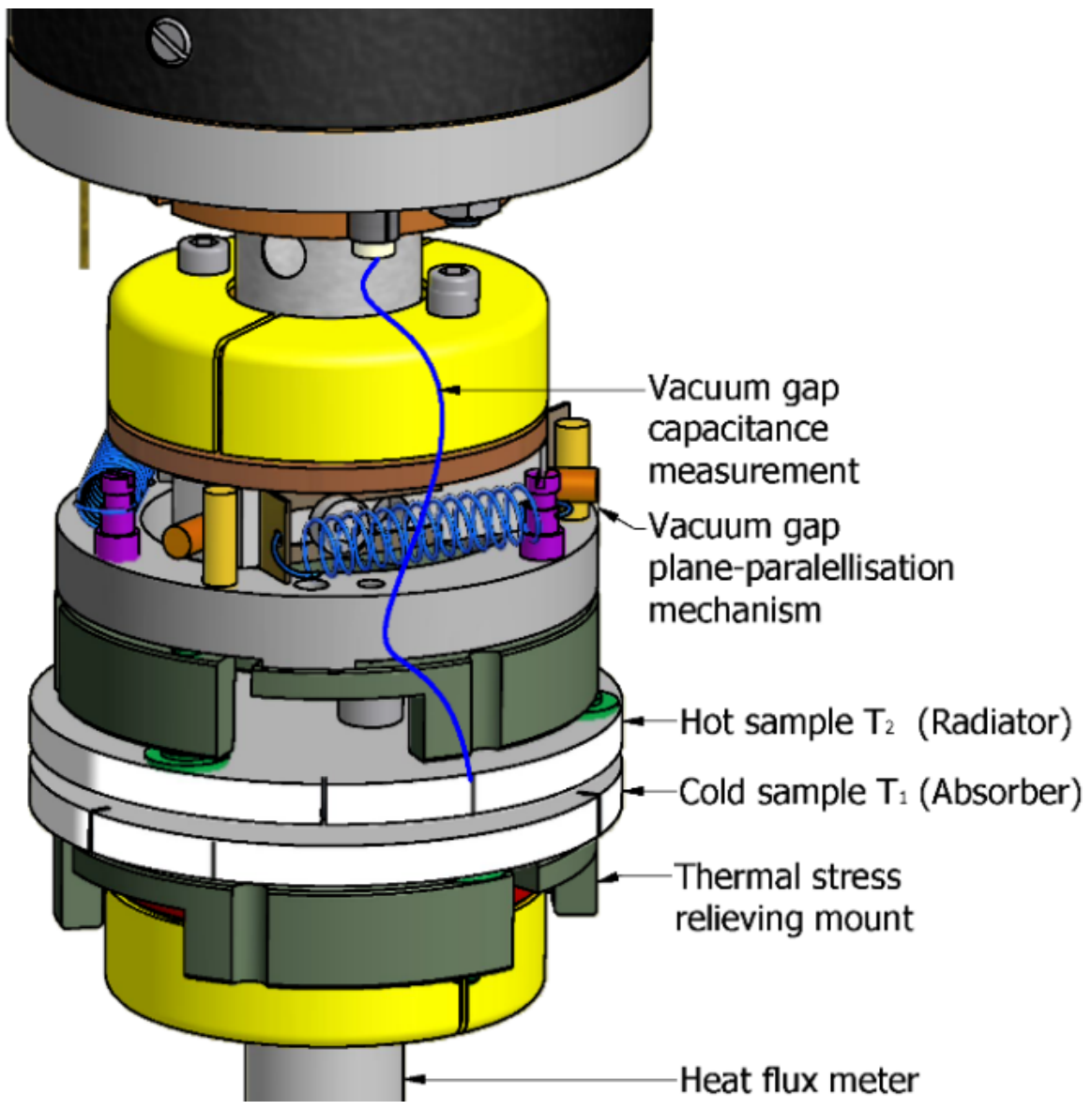

Fig. 2. (Color online) Scheme of the EWA measurement chamber core, containing the cold (absorber at temperature $T_1$) and hot (radiator at temperature $T_2$) samples with a mechanism for plane-parallelism control and a mount limiting the samples' planarity deformation due to thermal stress. The inter-sample distance *d* (i.e. the "vacuum gap") is determined via capacitance measurement. Adapted from Ref. [5].

A significant source of uncertainty in the measured power $P$ is due to external electromagnetic fields. Although the apparatus wiring is protected with inductor-capacitor filters, the external fields nevertheless cause parasitic heating of the absorber and of the HFM to a certain extent. This background can be measured and subtracted from the value of $P$. It can be done effectively when the background is stable, conveniently, e.g., during calibration thanks to the short time necessary for the measurement. On the other hand, the period of heat transfer measurement needs a much longer time to achieve steady state. We accept the measurements, in which the background fluctuations of the absorber temperature do not exceed the temperature resolution markedly and we estimate the effect of background from the reproducibility of power $P$.

The overall estimated uncertainty is caused by (i) the temperature resolution giving the lowest measurable power value $P$=0.1 μW with uncertainty of a few percent, which decreases with $P$ increasing; (ii) the unstable background (estimated from $P$-reproducibility), which is between 1 % and 10 % for $P$>0.5 μW ($q$>5×10$^{-4}$ W/m$^2$) and may be above 10 % for $P$<0.5 μW ($q$<5×10$^{-4}$ W/m$^2$) in some cases; and (iii) the uncertainty in the calibration curve is estimated as $\delta P$=0.02$P$.

We take an upper estimate for the total uncertainty of the measured power as

$$\delta P < 3.8 T_1^{1.222}(T_{1,res} + T_{0,res}) + (b+c)P, \quad (13a)$$

$$\delta P > 0.01 \quad [\mu\mathrm{W; K}], \quad (13b)$$

where the contribution due to the unstable background $b$ ranges between 0.01 and 0.1 and the one due to the calibration is $c$=0.02.

In EWA, the plane-parallelism of the samples is adjusted in situ by a special mechanism [5] and thereafter the sample spacing $d$ is set. At the highest values of $P$ (which is the case of near-field transfer over short distances), the uncertainty in $d$ has a great influence on the uncertainty in difference between the measured and theoretical value of $P(d)$, since

$$\delta P \approx -nP\delta d / d \quad (13c)$$

where $n$ is a power exponent of the theoretical dependence $P \sim d^{-n}$.

## IV. OVERVIEW OF RESULTS

In Figs. 3-5, we show three types of experiments, with a subset of the data, that provide a clear demonstration of the effect from various perspectives using the directly measured quantities: The transferred radiative power $P$ depending on (i) the absorber temperature $T_1$ at fixed [$T_2$, $d$], see Fig. 3; (ii) the spacing of the plates at selected fixed [$T_1$, $T_2$], see Fig. 4; and finally (iii) the radiator temperature $T_2$ at fixed [$T_1$, $d$]; see Fig. 5.

Figure 3 shows the power $P$ as a function of $T_1$, in the range between 5 and 17 K (the absorber undergoes the superconducting transition at $T_c$=15.3 K), at selected values of $d$ and $T_2$>$T_c$ (the radiator remains in the normal state). We can see a steep decrease of the NF power $P$ with the absorber temperature $T_1$ decreasing below $T_c$, while $T_2$ remains fixed well above the $T_c$. At the lowest temperatures of the absorber shown, the transferred power $P$ tends to saturate towards a nonzero value, which depends on $d$ and $T_2$.

Figure 4 shows $P$ instead as a function of the distance $d$ ranging from $d$~300 μm down to a few micrometers at temperatures held constant. Three values of $T_1$=5 K, 9 K, and 15.3 K (the latter just above $T_c$; cf. Fig. 1) at a single value of $T_2$=20 K are shown. We can see that the expected strong dependence of transferred radiative power $P$ on the intersample distance $d$ in the NF regime ($d$<50 μm in case shown), observed by us in [1, 5], is seen here for both NbN samples in the normal state but also when the absorber becomes superconducting for $T_1$<$T_c$. Entering the FF region at $d$>100 μm, the heat flux does not depend on the distance $d$ any more, in agreement with the observations in Refs. [1, 5]. The decline of power P seen here at $d$≈60 μm (at the crossover between the NF and FF regions) and for $T_1$<$T_c$<$T_2$, has not yet been observed in our measurements with Nb [1]. Below we argue that it occurs due to interference.

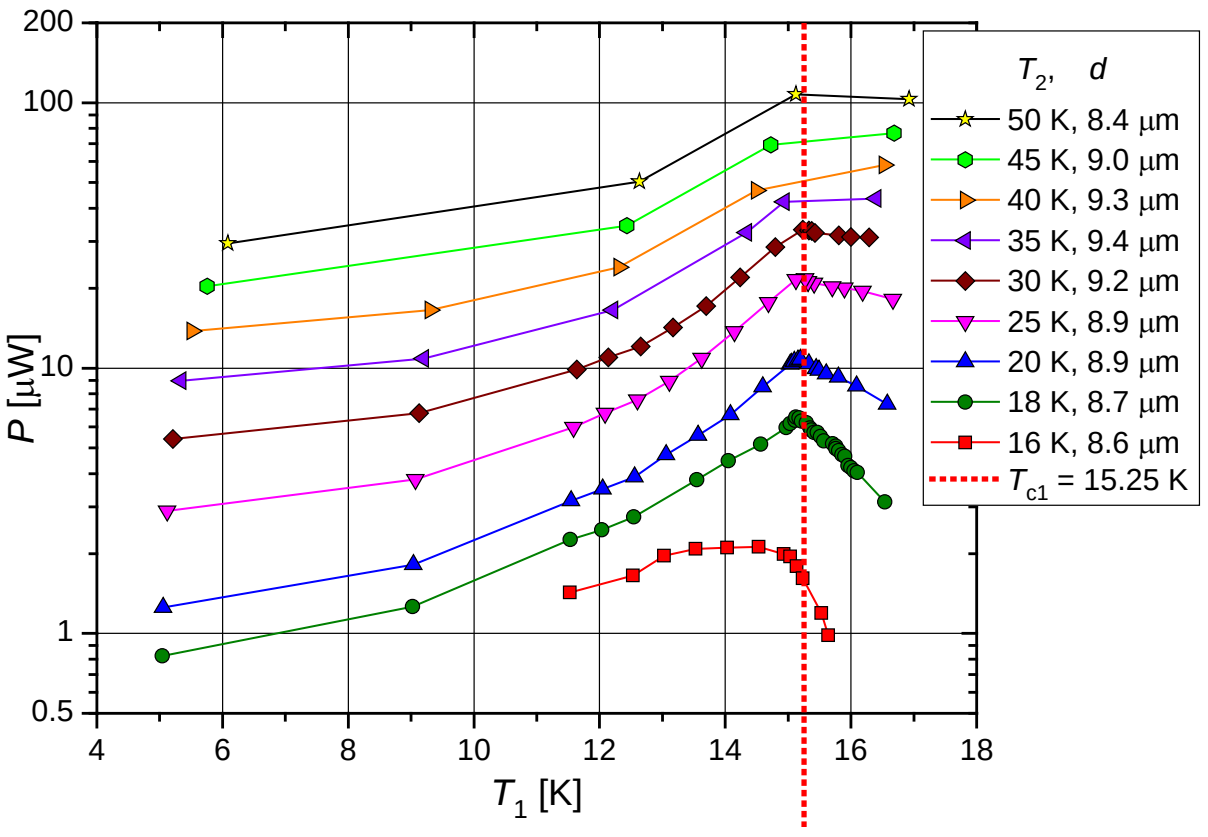

Fig. 3. (Color online) Measured radiative heat power $P$ transferred over the $d$≈9 μm wide vacuum gap between the plane-parallel NbN films, dependent on the absorber temperature $T_1$. The absorber passes from normal to the SC state at $T_1$=$T_{c1}$=15.25 K, as seen for several radiator temperatures $T_2$>$T_c$ between 16 K and 50 K.

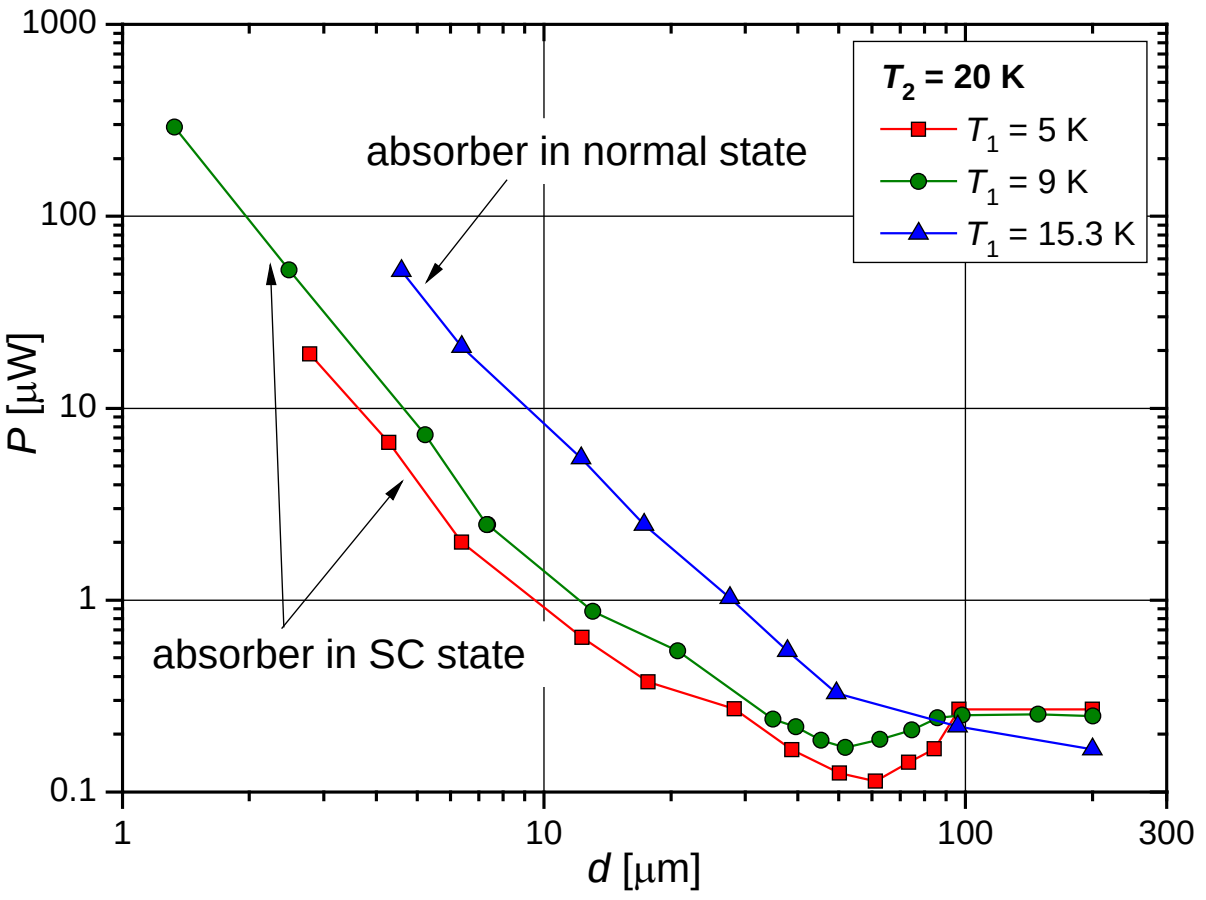

Fig. 4. (Color online) Measured radiative heat power $P$ dependent on the vacuum gap size $d$ between the plane-parallel NbN films. The radiator is in the normal state at $T_2$=20 K. The absorber is at temperatures $T_1$=5 K, 9.8 K (SC state) and 15.3 K (normal state). Note to the point at $d$ = 200 μm, $T_1$ = 15.3 K: data points $P$ < 0.5 μW might be affected by random background changes leading to errors exceeding 10% (see Sec. III. B).

Figure 5 shows $P$ as a function of the radiator temperature $T_2$ in the range between 14 and 26 K, at a fixed distance $d$=12.3 μm. The radiator undergoes the SC transition at $T_2$=$T_{c2}$=15.21 K, while the absorber is in the SC state at $T_1$=9.8 K. At least due to the limited resolution of the P measurement, we cannot distinguish the theoretically expected change in the slope (seen in a log-log plot, Fig. 5) at the $T_{c2}$ transition. Similarly, the apparent waviness near the SC transition cannot be distinguished within the present resolution as a real effect. We do not discuss the experiment with both samples in the SC state further here. Note only that the measured heat power is higher than the theoretical one by a factor ~1.5 which increases with decreasing temperature $T_2$ up to ~2.5 at $T_2$=14 K.

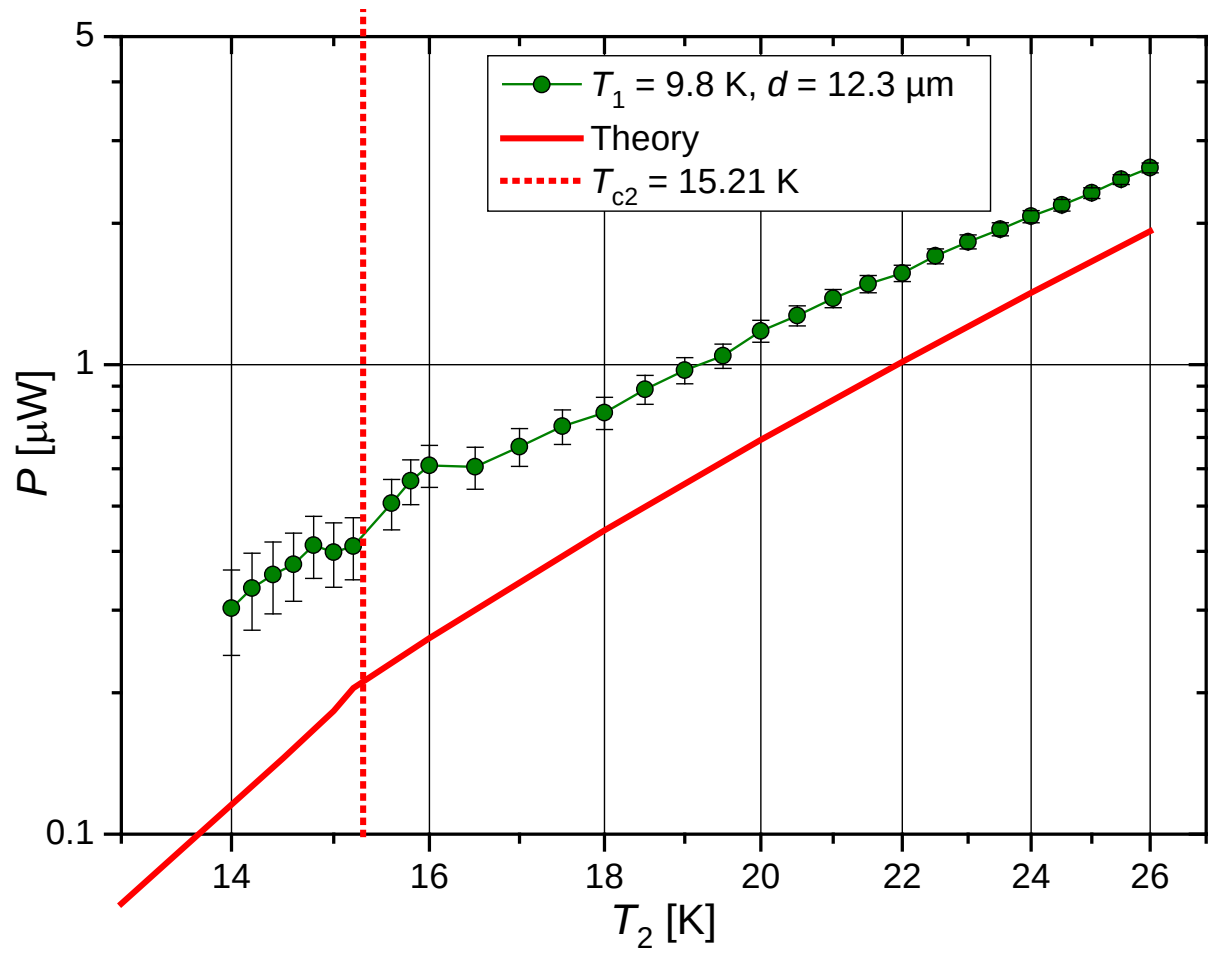

Fig. 5. (Color online) Measured radiative heat power transferred over the vacuum gap dependent on the radiator temperature $T_2$ passing the SC transition at $T_2$=$T_{c2}$= 15.21 K. The absorber is in SC state at constant $T_1$=9.8 K. The error bars show the *P* resolution derived from the temperature resolution of 50 µK, Eq.13

## V. DISCUSSION OF RESULTS

### A. Heat transfer limits

Figure 6 shows the dependences of the heat flux *q=P/A*, where *A* is the samples area, on the temperature difference $T_2$ - $T_1$ for three distances *d* in the near field regime. The data can be understood as a replot of Fig. 3 (with extensive data added) showing the dependences on $T_2$ - $T_1$ for three different distances *d*≈5.4 µm (a), *d*≈9.0 µm (b), and *d*≈12.4 µm (c), for a set of radiator temperatures $T_2$ from 16 to 50 K. Notice that the two theoretical curves corresponding to $T_1$=5 K and 15.3 K create "approximate envelopes" to the data. The upper envelope corresponds to both samples in the normal state at a constant absorber temperature $T_1$=15.3 K, while for the lower envelope, the absorber is in the SC state at $T_1$=5 K. The data between the envelopes (region of negative thermal conductivity of the vacuum gap) are obtained with absorber in the SC states. The theoretical curves of q($T_2$-$T_1$) calculated for constant values of $T_2$ show that decreasing the temperature $T_1$ below 5 K down to the absolute zero means only a negligible decrease in the heat flux. The calculated envelopes present two (approximate) power-law limits of the heat flux dependences, namely, (i) q ~ $(T_2 - T_1)^m$, m≈1, for both samples in the normal state at $T_1$ and $T_2$ above $T_c$=15.25 K, and (ii) the same, but with m≈3, corresponding to a SC absorber approaching $T_1$≈0 K.

The ratio of heat flux values at the two limits

$$C\left(T_2,d\right)=q\left(T_1=T_c,T_2,d\right)/q\left(d,T_1=5K,T_2,d\right) \quad (14)$$

is conveniently characterized as the "contrast," following Ref. [1]. Figure 7 plots the dependence of the theoretical contrast *C* (derived from the upper and lower envelope curves in Fig. 6) on the radiator temperature $T_2$ for the three distances *d* considered. The maximum of *C* is seen near $T_2$=20 K. Interestingly, the effect of superconductivity on the NF heat transfer is seen even at $T_2$≈4$T_c$=60 K; i.e., it persists well above the radiator critical temperature. We confirmed this effect experimentally up to $T_2$=50 K, where the contrast reaches *C*≈4 at *d*=8.4 µm. The maximum experimental value, *C*≈8, was observed at $T_2$=20-25 K and *d*=5.5 µm.

### B. Temperature dependences

For comparison of the temperature dependences above and below the absorber superconducting transition, we introduce the vacuum gap conductivity:

$$K=q/\left(T_2-T_1\right). \quad (15)$$

The choice is inspired by the upper curve in Fig. 6, following nearly a linear dependence $(T_2 - T_1)^m$ with m≈1, which means a constant vacuum gap conductivity $K(T_1,T_2)$ when both $T_1$ and $T_2$>$T_c$.

This follows from two facts: First, in the NF regime, the vacuum gap transmissivity between normal metals is extremely high at long wavelengths, for which the

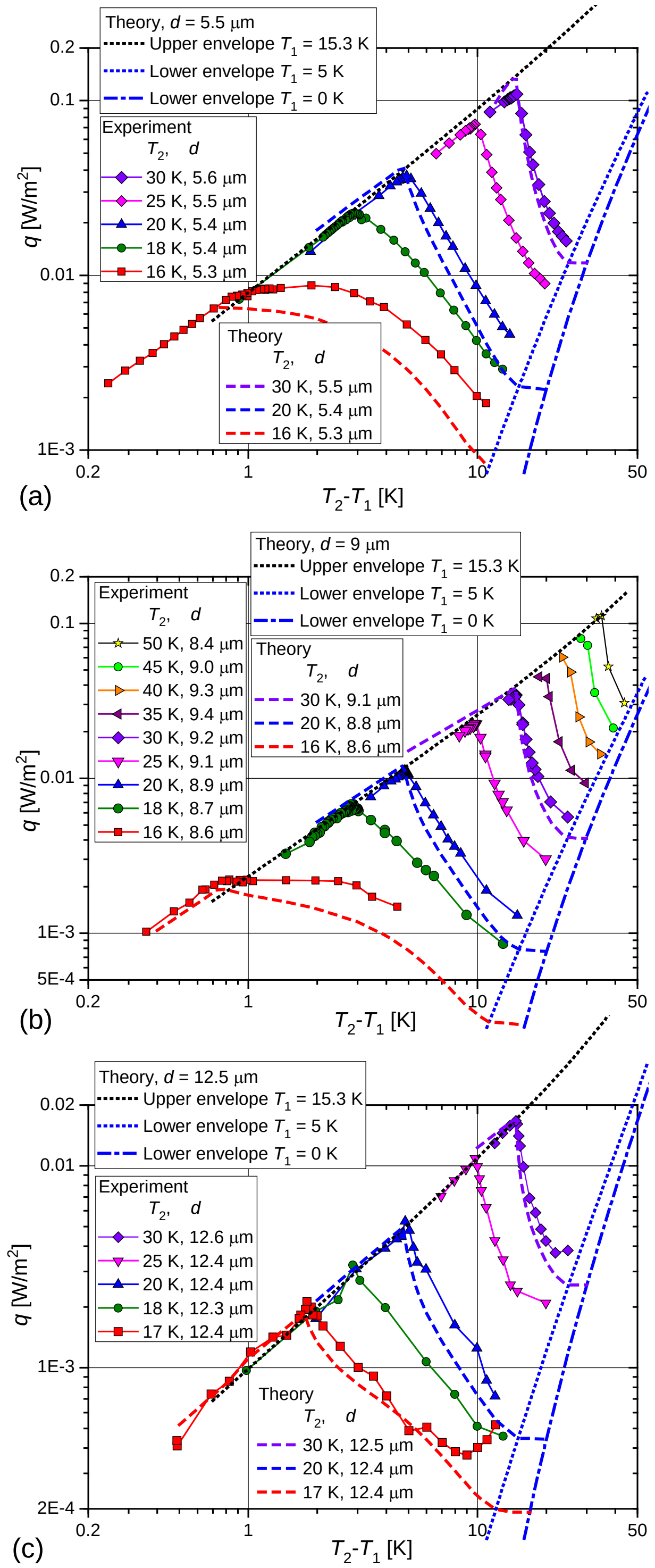

Fig. 6. (Color online). Heat flux dependent on the temperature difference $T_2$-$T_1$ between the radiator and absorber measured for a set of radiator temperatures $T_2$ >$T_c$ for vacuum gap widths $d$≈5.5 µm (panel a), ≈9 µm (b) and ≈12.4 µm (c). At a constant radiator temperature $T_2$>$T_c$, the absorber state was varied from normal metallic to superconducting (data points from left to right in the plot). Dashed lines plot theoretical dependences for selected temperatures $T_2$. Theoretical "envelopes" of the data correspond to varying $T_2$ at constant $T_1$≈$T_c$=15.25 K (upper envelope), $T_1$=5 K lower envelope) and $T_1$=0 K (dot-dashed line).

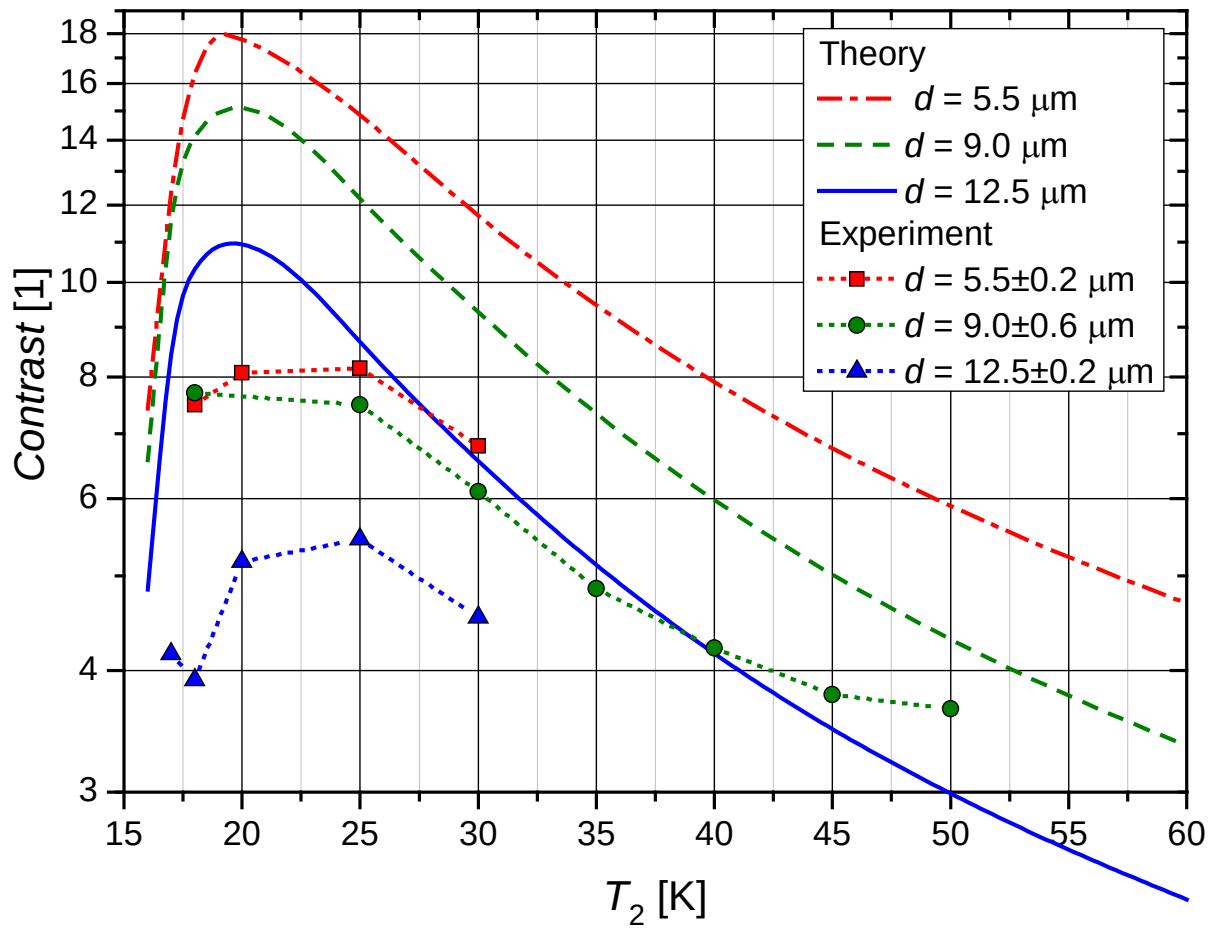

Fig.7. (Color online) The measured (data points) and theoretical (solid lines) contrast $C$, Eq. (14), in the NF regime dependent on the radiator temperature $T_2$ at three distances d between the NbN layers.

radiation intensity given by Eqs. (1) and (2) simplifies to

$$I(T_1,T_2,\omega) \approx \frac{1}{\pi}\left(\frac{\omega}{2\pi c}\right)^2 k_B(T_2-T_1), \quad \omega \to 0. \tag{16}$$

Second, the vacuum gap transmissivity between the NbN metallic films is nearly constant at temperatures $T_1$, $T_2$>$T_c$ due to the weak temperature dependence of the electrical (and optical) properties of NbN in the normal state (see Fig. 1).

In Fig. 8, we replot the data of Fig. 6 as a dependence of the conductivity $K$ on $T_1$ for various radiator temperatures $T_2$ together with the theoretical curves. We observe a steep decrease in $K$ for $T_1$<$T_c$ down to half the normal-state value at about $T_1$≈0.9$T_c$ (experiment) and at 0.90$T_c$<$T_1$<0.95$T_c$ (theory). The decrease saturates at low temperatures.

Above the critical absorber temperature $T_{c1}$, the experimental data in Fig. 8 agree with the theoretical model within 10 %-20 % uncertainties, which correspond to the possible uncertainties in the distance $d$; see Eq. (13c). For example, in a measurement at $d$=5.5 µm, the uncertainty of 0.25 µm in $d$ would cause an uncertainty of about 10% in the NF heat flux in the normal state, where the heat power $P$~1/$d^n$, with $n$≈2.5 (cf. Fig. 4).

On the other hand, the low-temperature heat flux below $T_{c1}$ is systematically higher than the theoretical values by a factor of about 1.5-2. This corresponds to a less steep decrease of the measured heat flux following the superconducting transition compared with the theory; cf. the contrast shown in Fig. 7.

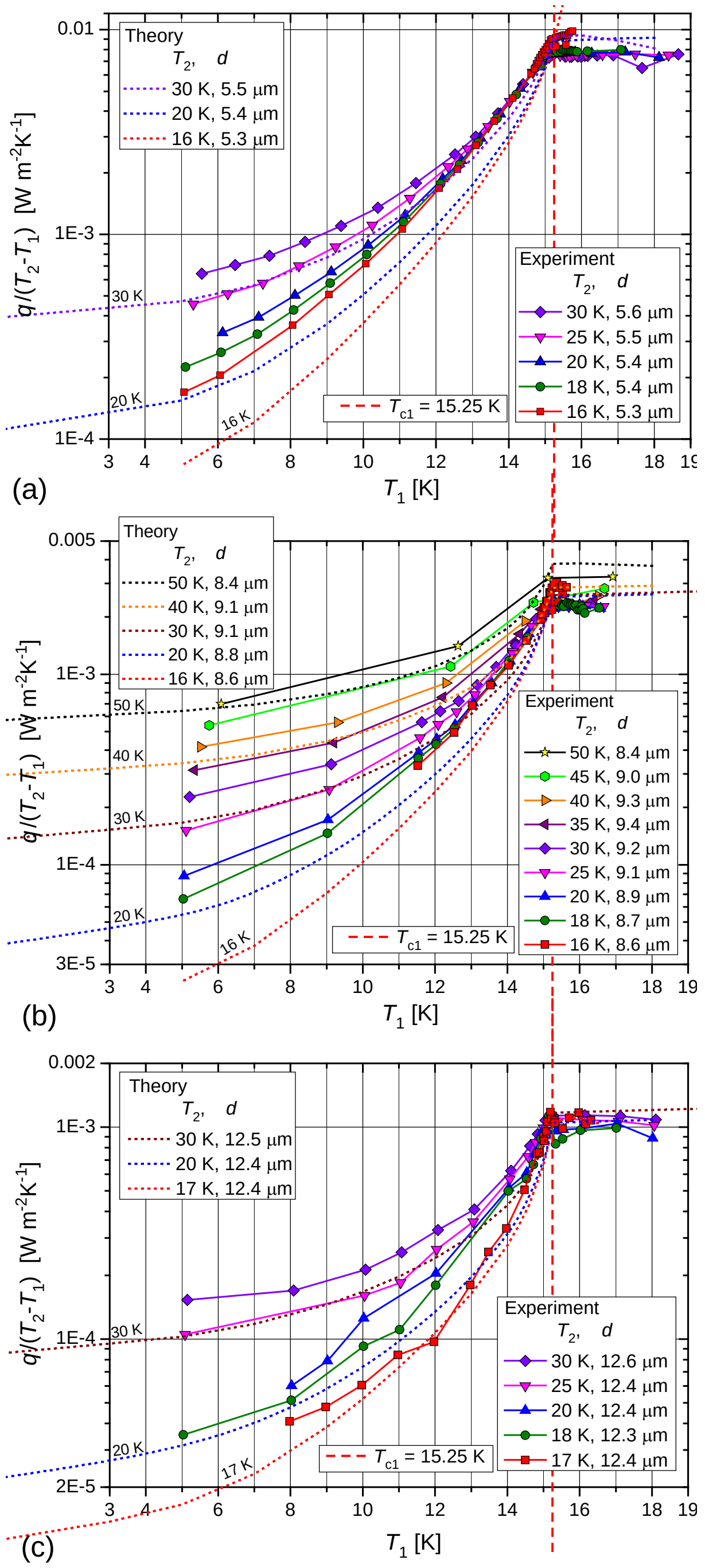

Fig. 8. (Color online) Thermal conductivity $K=q/(T_2-T_1)$ of the vacuum gap dependent on the absorber temperature $T_1$ measured at a set of fixed radiator temperatures $T_2$ for three vacuum gap widths: $d$≈5.5 µm (panel a), $d$≈9 µm (b), and $d$≈12.4 µm (c).

### C. Distance dependences

A useful variable to compare the distance dependences of the NF heat flux is the mutual emissivity of the samples

$$e = q / q_{bb}, \qquad q_{bb} = \sigma_{SB}(T_2^4 - T_1^4), \tag{17}$$

defined as the measured heat flux normalized to the classical FF heat flux between black surfaces, where $\sigma_{SB}$ denotes the Stefan-Boltzmann constant.

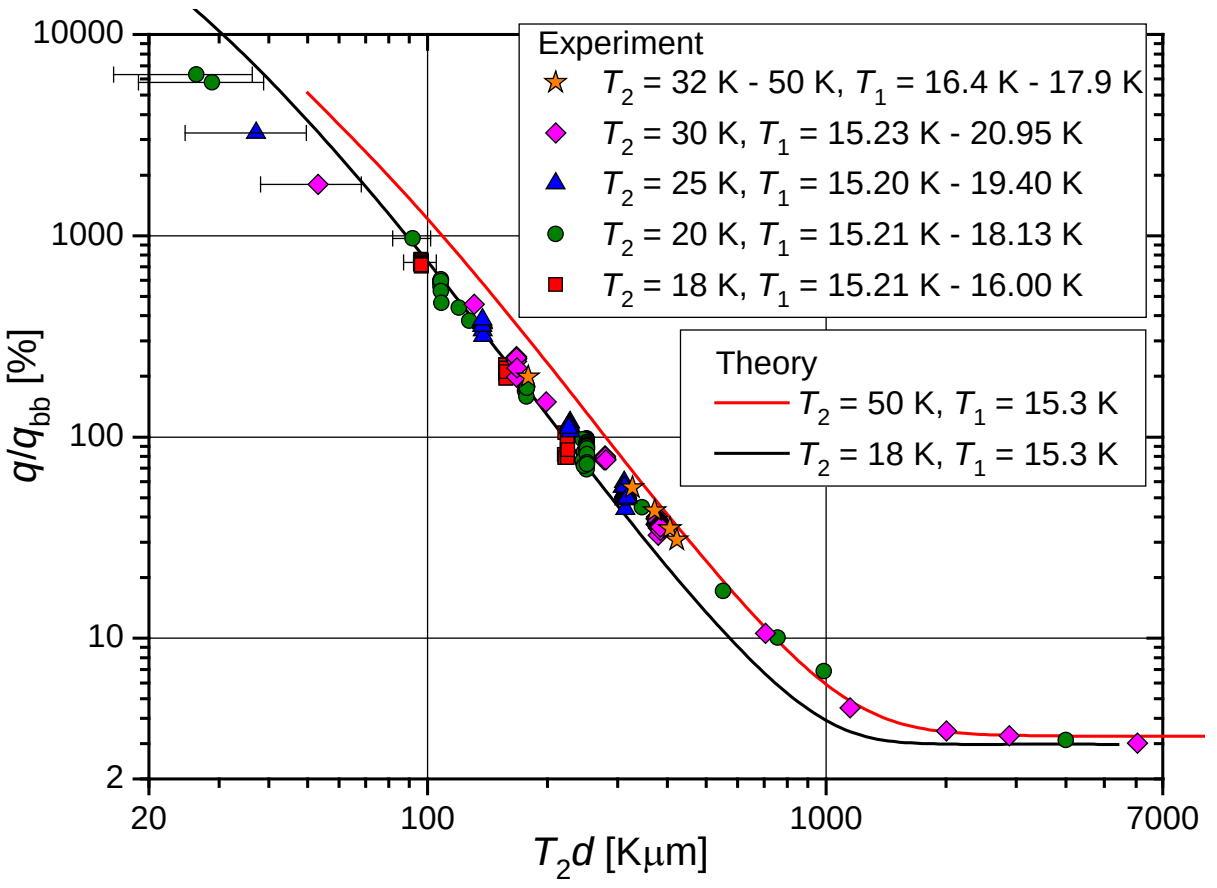

Fig. 9. (Color online). The mutual emissivity (heat flux $q/q_{bb}$ "normalized to the blackbody", Eq. 17) dependent on the product $T_2d$ for both samples in normal state. Data are collected from measurements of $q(d)$ and $q(T_1)$ dependences, where the remaining two parameters $T_1$, $T_2$ and $T_2$, $d$, respectively, are fixed. Error bars show the uncertainty in $T_2d$ corresponding to inter-sample distance $d$ uncertainty of 0.5 µm [cf. Eq. (13), last term].

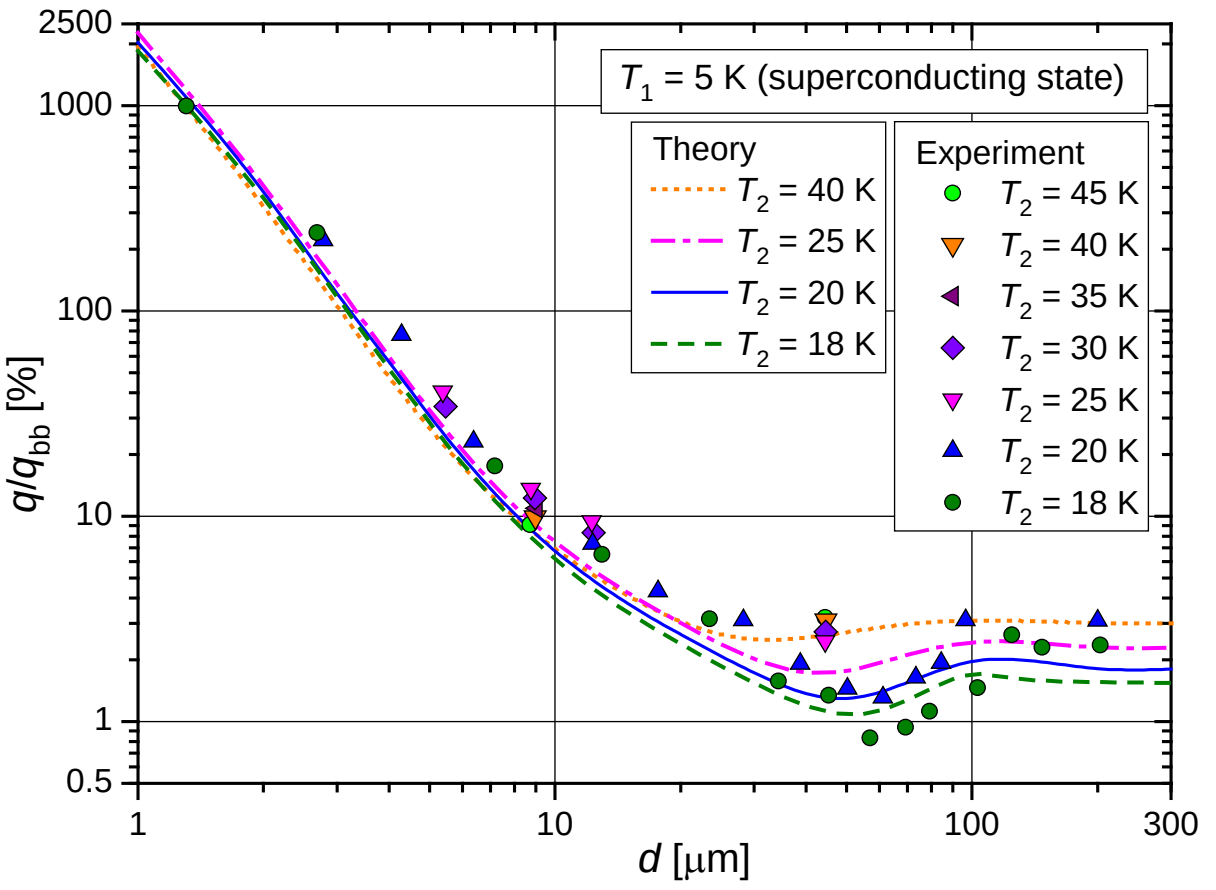

Fig. 10. (Color online) The same as in Fig. 9 but depending on d instead of $T_2d$, for the absorber at $T_1$=5 K (superconducting) while the radiator is in normal state at various temperatures $T_2$ ranging 18 K–45 K.

For both samples in the normal state, the linear dependence of the NF heat flux on ($T_2$ - $T_1$), together with the observed dependence on $d$ (cf. Fig.4),

$$q \sim const.(T_2 - T_1)/d^n, \qquad 2 < n < 3, \tag{18}$$

lead to the following relation for the emissivity

$$e \approx \frac{const.(T_2 - T_1)}{\sigma_{SB}(T_2^4 - T_1^4)} d^{-n} \approx \frac{const}{T_2^{(p-n)}} (T_2 d)^{-n}. \tag{19}$$

When we take into account that in our region of temperatures $(T_2^4 - T_1^4) \approx (T_2 - T_1)\, T_2^{\,p}$, $p \approx 3$, and that $p$-$n$<1, we see that the term $T_2^{p-n}$ varies only weakly with $T_2$. This is demonstrated in Fig. 9, where the theoretical lines corresponding to $T_2$=18 K and 50 K follow one another very closely. Although the experimental values of the emissivity measured at distances between 5 μm and 15 μm agree well with the theoretical values, the experimental points obtained over three decades of heat flux values follow a slightly flatter dependence on $T_2d$ compared to the theoretical curves (cf. also Fig. 13). Let us note here, that we observed a similar flattening previously for tungsten samples [5].

On the other hand, with the absorber in the SC state, the theoretical curve $q(T_2\text{-}T_1,\ T_1\text{=}5\ \text{K})$, lower envelope in Fig. 6, follows approximately the third power law, which approaches the blackbody power law for heat flux at corresponding temperatures. In this case, the NF emissivity, Eq. (17), varies slowly at $T_1$=5 K when $T_2$ varies at a constant *d*. Therefore, the NF emissivity data obtained for various $T_2$ follow nearly the same *d*-dependence (Fig. 10). Notice the difference from Fig. 9, where a "data collapse" occurs dependent on the product $T_2d$. As observed already in the $T_1$–dependences shown in Fig. 8, the points obtained in the NF regime (*d*=2.5-30 μm) with the superconducting absorber are systematically higher by a factor of 1.5-2 compared to the theoretical values.

### D. Roles of TE and TM modes

To understand more deeply the experimental and theoretical *T* (Figs. 3, 6, and 8) and *d* (Figs. 4, 9, and 10) dependences of the radiatively transferred heat, we analyze here the detailed behavior of individual TE, TM modes of both the NF and FF components obtained by solving Eq. (3) within the sample model in Sec. II C. Figure 11 shows the TE and TM components of the NF modified spectral transmissivity *M*, Eq. (7), for the vacuum gaps *d*=5.4 μm (a) and *d*=12.4 μm (b) wide.

For both samples in the normal state, $T_1$>15.25 K, the transmissivity *M* does not depend on the temperatures $T_1$, $T_2$ due to the independence of the NbN optical properties for frequencies and temperatures in our range of interest. At long wavelengths (low *ω*) in the NF regime, the TE mode strongly dominates the TM mode which is (in contrast to dielectrics) typical for metals [3], [10]. The spectrum of the dominating TE mode is broader for the shorter distance, *d*=5.4 μm, as consistent with the general condition of *λ*>*d* for wavelengths to be substantial in the NF effect. The $M_{TM}$ and $M_{TE}$ dependences cross each other, so that $M_{TM}>M_{TE}$ at high frequencies. For *d*=5.4 μm and *d*=12.4 μm this crossover is near above and just at the edge of the NbN SC energy gap $\omega_{g0} = 7.81\times10^{12}$ rad/s, respectively. Thus in the (radiator) normal state, the major contribution of the TE mode falls into the (absorber) superconducting gap $\omega_{g0}$. Pronounced dominance of the TE over the TM mode is seen at the distance *d*=5.4 μm [Fig. 11(a)], while the dominance is weaker (although still strong) for d=12.4 μm [Fig. 11(b)].

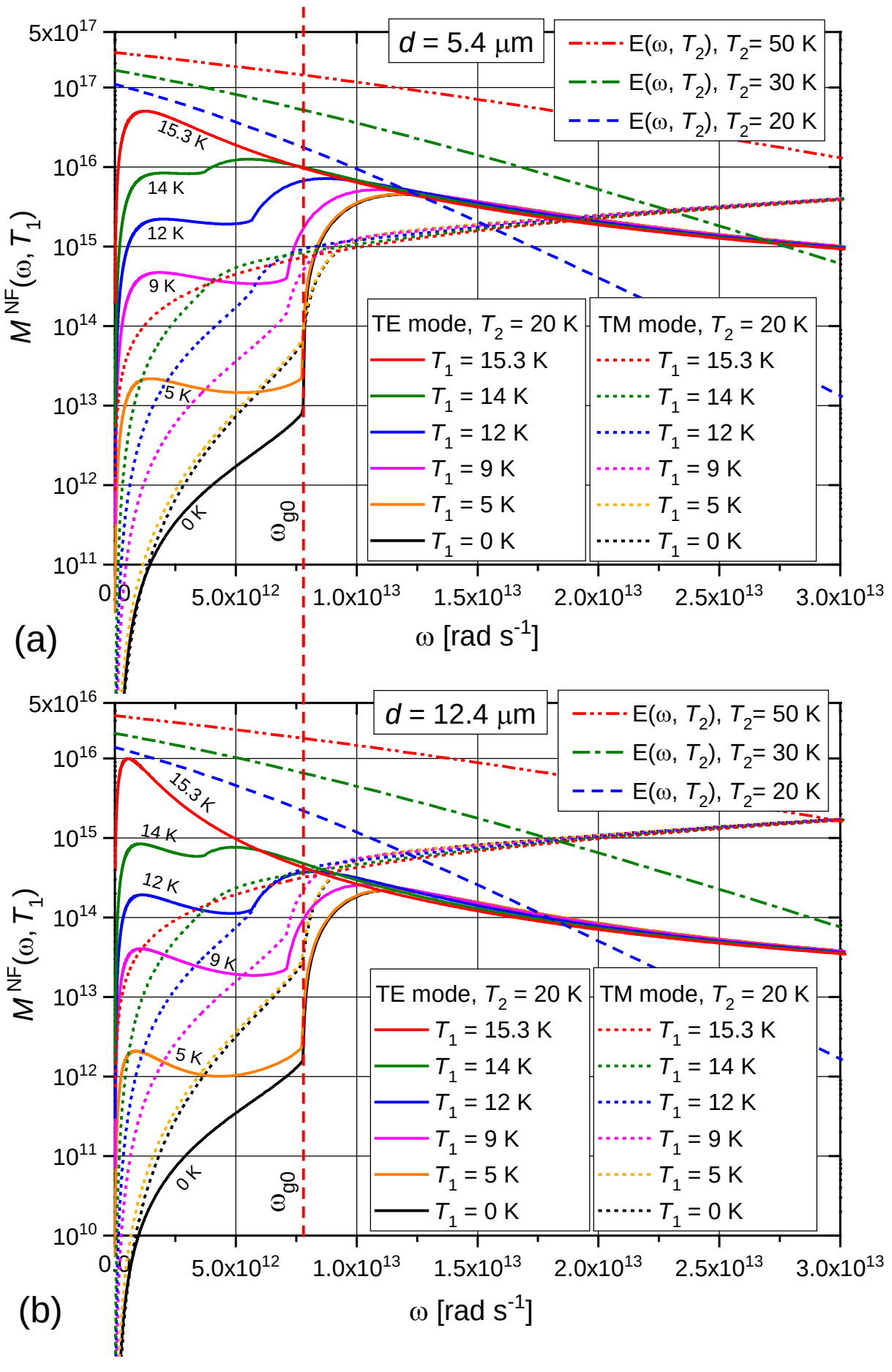

Fig. 11. (Color online) Spectrum of modified vacuum-gap NF transmissivity $M^{NF}$, Eq. 7, dependent on the transferred heat flux frequency *ω* at various absorber temperatures $T_1$ for *d*=5.4 μm (panel a) and *d*=12.4 μm (b). Notice that the TE and TM components of $M^{NF}$ are shown for radiator temperatures $T_2>T_c$, where M practically does not depend on $T_2$. For comparison, we show also the (monotonous) oscillator energy *E* dependences (multiplied by a constant) for three radiator temperatures.

Superconductivity of the absorber, $T_1$<15.25 K, suppresses the theoretical transmissivity within the SC energy gap (Fig. 11), causing a steep decrease in heat flux with decreasing absorber temperature $T_1$ after crossing $T_{c1}$ (cf. Fig. 8).

Using transmissivities *M* plotted in Fig. 11 for solving Eq. (5), we obtain heat flux and its TE, TM components. For the two respective distances, they are plotted in Figs. 12(a), and 12(b) dependent on $T_1$. At a distance of 5.4 μm the TE mode keeps domination over the TM mode even in the SC state $T_1<T_c$ [Fig. 12(a)] while at the longer distance *d*=12.4 μm and at temperatures $T_1$<11 K, the TE mode is suppressed below the TM one. Suppression of the TE mode by superconductivity thus "denudes" the TM mode contribution at longer distances *d* in the NF regime, which is also seen in *d*-dependences, where this dependence of heat flux (emissivity) is less steep, ~1/*d* at d>10 μm (Fig. 10), than the dependence where the TE mode dominates. For $d$ dependencies of theoretical curves, see Figs. 13 and 14 in the next paragraph where strong suppression of the NF TE mode is evident.

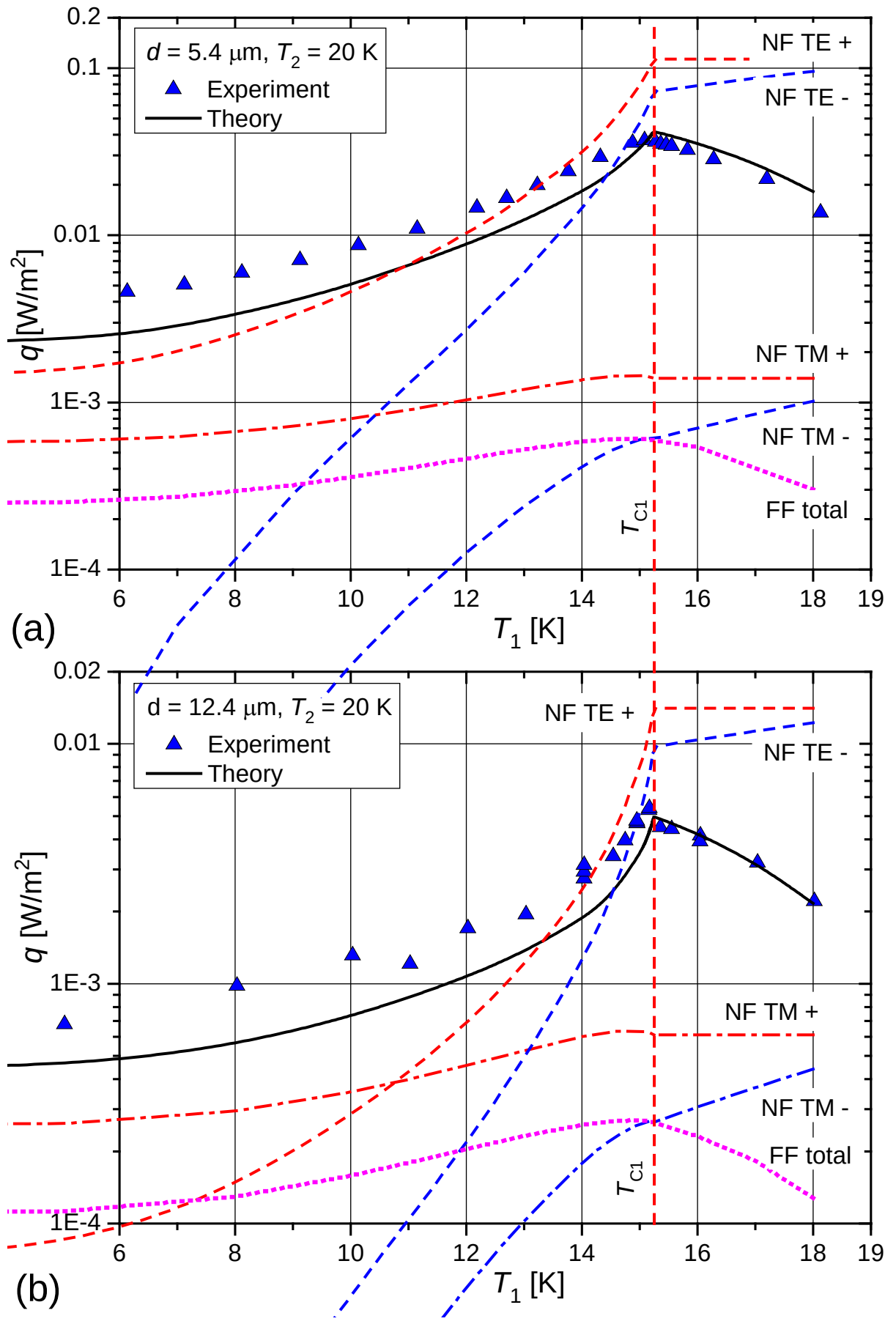

Fig. 12. (Color online) Heat flux $q$ over the vacuum gap $d$=5.4 µm (panel a) and $d$=12.4 µm (b) between the NbN sample layers (experiment-points, theory-lines). The radiator is in normal state at $T_2$=20 K, while the absorber undergoes the SC transition at $T_{c1}$=15.25 K. NF heat fluxes from the radiator at $T_2$ to absorber at $T_1$ (labeled with "+") and vice versa ("-") are distinguished (their difference gives the total NF heat flux). Notice that the FF total heat flux (dotted line) is small at these inter-sample distances $d$ throughout the $T$-range considered.

### E. Interference of heat radiation

Figures 13 and 14 show the theoretical contributions of the TE and TM components of the total heat flux as a function of $T_2 \cdot d$, corresponding to the absorber in normal and SC states, respectively. In both cases, the TM mode is dominating in the FF regime, while in contrast, the NF regime shows a crossover: TE dominates at short distances, while TM dominates at long distances.

In Fig. 14, we see that when the absorber is in the SC state at $T_1$=9 K, the total heat flux decreases to a minimum near a distance $d$ given by $T_2 d$≈1000 K µm ($d$≈50 µm at $T_2$=20 K; cf. also Fig. 10 where $T_1$=5 K). In contrast, the minimum does not appear for both samples in the normal state (Fig. 13).

Comparing Figs. 13 and 14 we can see that the dip in the heat flux in the SC state appears due to two concurrent effects: (i) deepening of the minimum in FF TM contribution and (ii) suppression of the NF TE mode, increasing thus the role of the FF TM mode in the total heat flux.

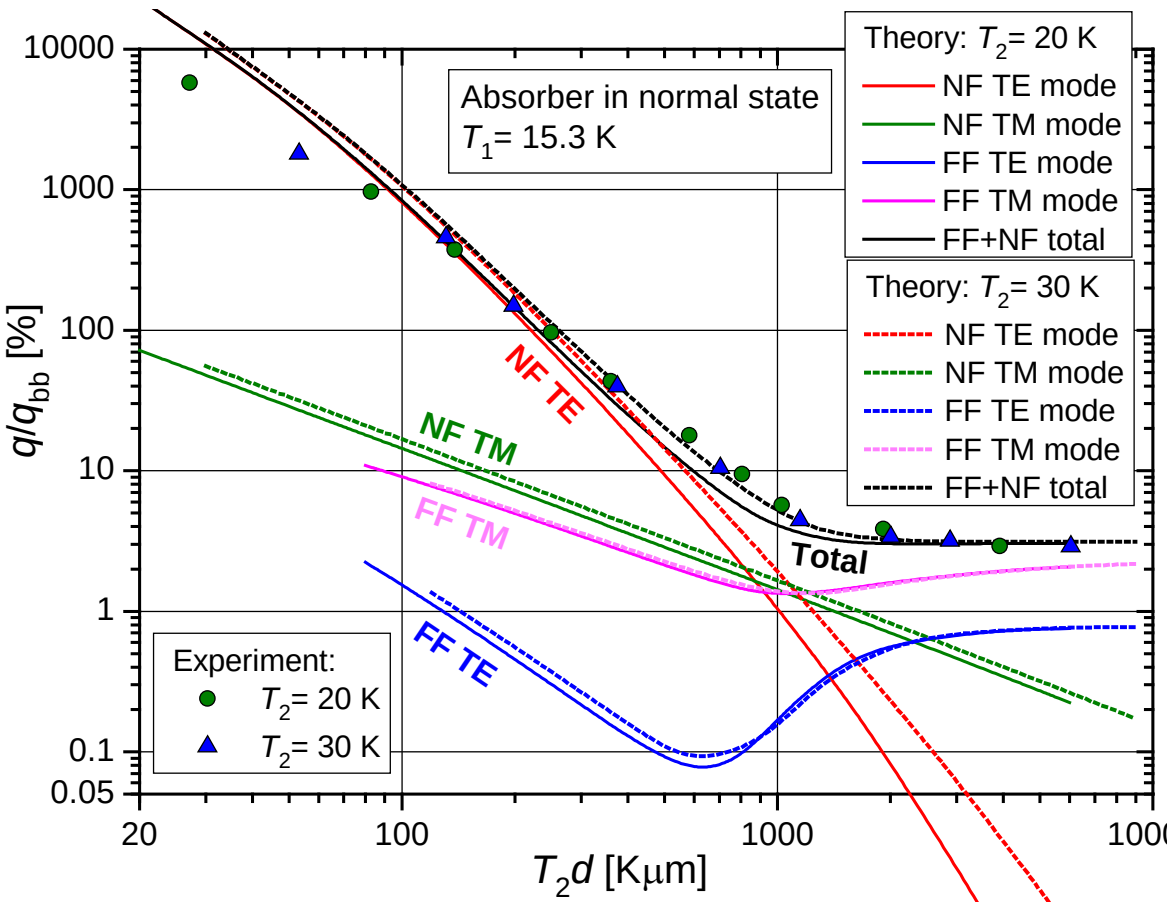

Fig. 13. (Color online) Theoretical (lines) and experimental (points) mutual emissivity $e=q/q_{BB}$ depending on the product $T_2 d$, for both samples normal. The radiator temperatures are $T_2$=20 K and 30 K, while the absorber is at $T_1$=15.3 K. The theoretical heat flux curve is decomposed to the TE and TM modes of both the NF and FF contributions.

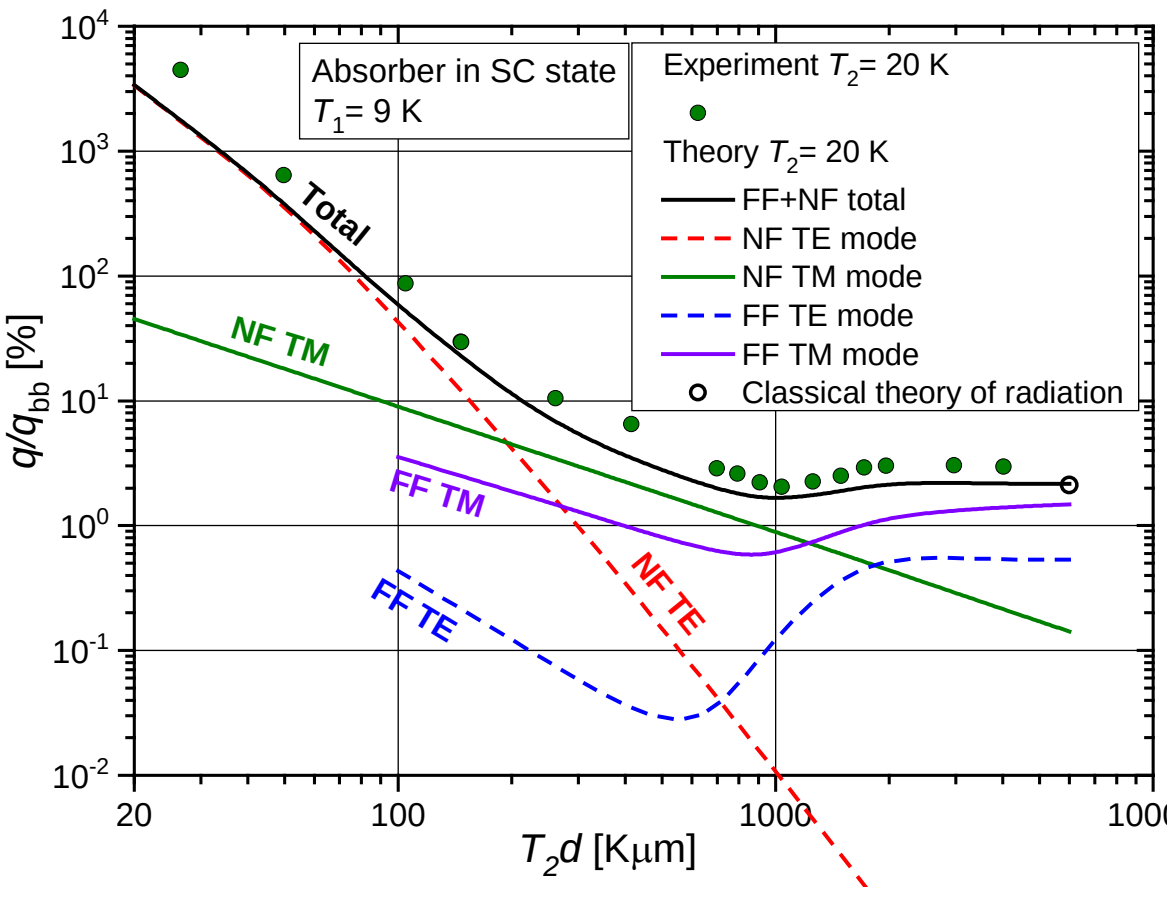

Fig. 14. (Color online) The same as in Fig. 13 for the radiator at $T_2$=20 K (normal) and absorber at $T_1$= 9 K (superconducting). The open circle point corresponds to the classical FF heat flux. The remaining points (full green circles) are measured.

The origin of the minimum in the FF TM contribution lies in interference [3] of the thermal radiation, as can be understood by comparing the spectra shown in Figs. 15(a) (normal-normal case; $T_2$ and $T_1$>$T_c$) and 15(b) (normal-SC case; $T_1$<$T_c$<$T_2$), which are plotted at three distances $d$=2000 µm, 200 µm, and 50 µm. At the largest distance $d$=2000 µm, the spectra contain numerous interference "wiggles," which only little perturb the far field heat transfer when neglecting the interference effect, setting $\exp(2i\gamma_0 d) \equiv 1$ in Eq.

(4a). They become more pronounced at shorter distances $d$, and for $T_2$=20 K cause a significant interference effect at $d$=50 µm. The effect of interference is stronger in the SC state as the radiation is “more monochromatic” compared to the normal state [Fig. 15(a)].

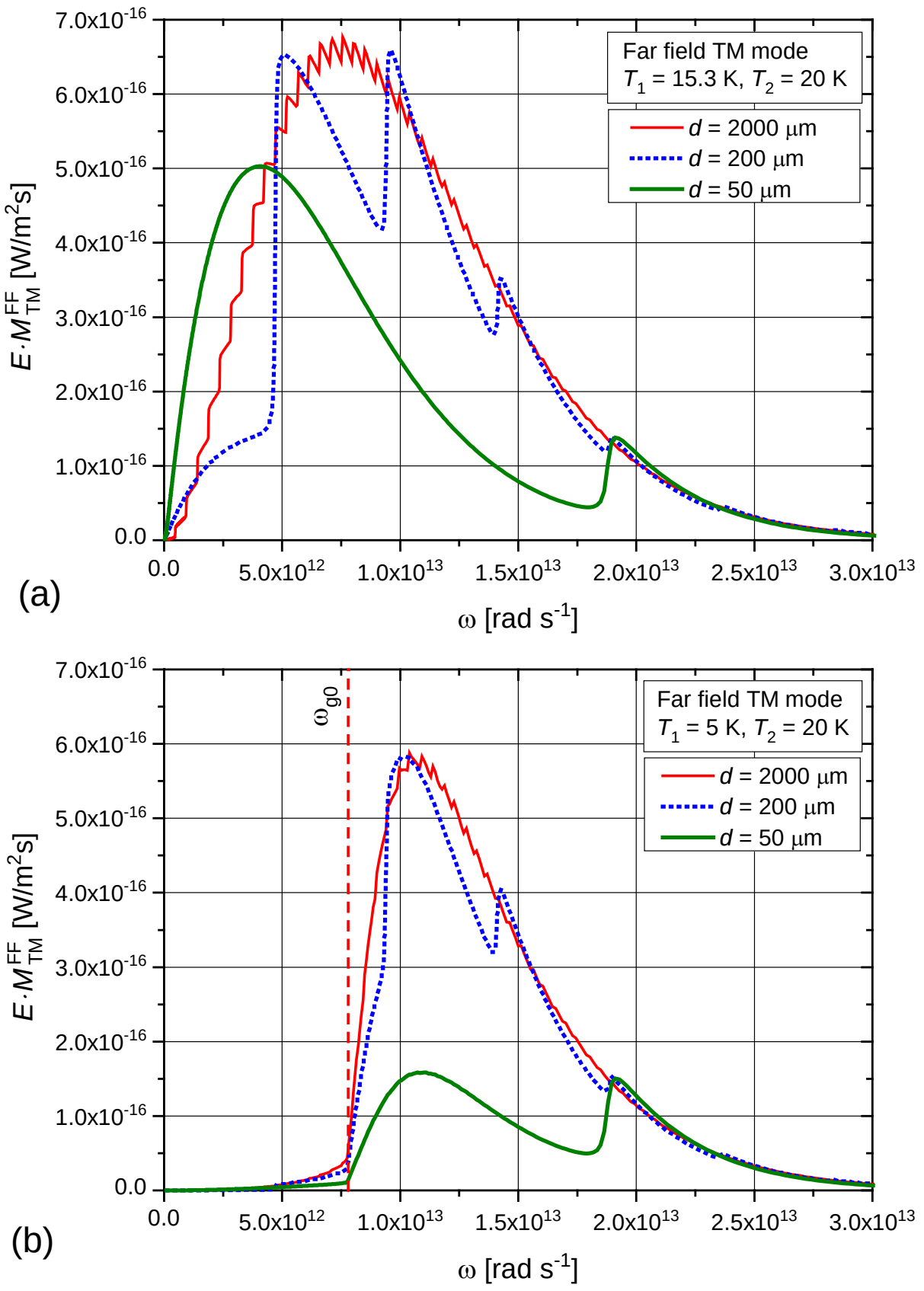

Fig. 15. (Color online) Full spectra E.M, Eqs. (6, 7), of the FF heat flux TM mode at a distance $d$=50 µm and 200 µm are compared with those corresponding to a large vacuum gap of $d$=2mm. The destructive interference apparent at $d$=50 µm is more effective in the SC state (panel b) in which the radiation is „more monochromatic“ due to the energy gap cut-off below $\omega_g \approx 8\times10^{12}$ rad/s. Wavelengths at maxima of FF TM spectral heat flux over vacuum gap d=2000 µm are λ≈250 µm (panel a) and λ≈160 µm (b).

The minimum in heat transfer has been expected for Nb in Ref. [1], where the theory predicts noticeable heat flux minima with superconducting as well as with normal absorbers. However, it could not be detected in the experiments with Nb in principle, due to low $T_c$ and low FF emissivity of Nb, giving the theoretically expected heat fluxes below the experimental resolution.

## VI. SUMMARY AND CONCLUSIONS

We performed a series of experiments on radiative heat transfer between plane-parallel thin-film NbN samples across the phase transition from normal-metal to superconducting states (at $T_c$=15.25 K). The samples were $L$=270 nm thin, and the vacuum gap $d$ between them was varied between ≈2 and 300 µm. Temperatures of the radiator varied between 14 and 50 K, while the absorber temperatures varied from 21 down to 5 K. We interpreted the experiments in terms of the Polder and Van Hove theory of heat transfer over vacuum gap [3] together with the Drude model and the BCS theory [27,28] for the respective sample conductivities.

The weak coupling BCS theory describes very well essentially all the experimentally observed features of the heat transfer dependences on the radiator and absorber temperatures $T_1$, $T_2$ and on the intersample distance $d$. A noticeable discrepancy occurs at the lowest temperatures of the SC absorber sample, where the heat transfer was systematically higher by a factor between 1.5 and 2. The discrepancy is with high probability not caused by uncertainties in the measured parameters of the samples that enter the theoretical model calculations, namely the critical temperature $T_c$, the direct current conductivity $\sigma_{dc}$, the sample thickness $L$, as well as the influence of the substrate, as we carefully verified (Supplemental Material [24]). We also cannot suspect the measurement method of heat flux and distance since the discrepancy factor occurs over more than two decades of those measured quantities.

We summarize the essential findings below:

1. *With both NbN samples in the normal state*: We see a nearly linear dependence of NF heat flux on the temperature difference $T_2$ - $T_1$ between the radiator and absorber surfaces, which reflects the long-wavelength character of the NF. The linear dependence can be generally expected for NF heat transfer in materials with weak $T$ dependence of the optical properties for $T$>$T_c$.
2. *With the absorber superconducting and the radiator normal*: We see a steep decrease of the heat flux at the transition to superconductivity for $T_1$<$T_{c1}$, leading to very high values of contrast C≈8. This differs dramatically from the situation in the FF regime, where only a weak effect of superconductivity was observed. Surprisingly, the contrast remains high even at high radiator temperatures (e.g., at $T_2$=50 K, the measured contrast was $C$ > 3), with the characteristic energy of photons far above the SC energy gap. This is caused by a suppression of the (otherwise strong) TE mode contribution at frequencies below the SC energy gap $\omega_g \equiv E_g / \hbar$. The TM mode contributes to the heat flux at higher frequencies in metals, and thus the TM contribution is less sensitive (relative to TE) to the absorber superconductivity at radiator temperatures $T_2$>$T_c$. With decreasing temperature $T_1$, both the TE and TM mode heat fluxes *saturate* at values transferred mainly at frequencies above the energy gap. A dip occurs in the total heat flux for SC absorbers at distances just above the NF-FF crossover, which was up to now not observed experimentally. It is due to a *destructive interference* of the FF TM mode.

3. Finally, *with both samples superconducting*: Due to limited resolution of measurement, we could not distinguish any details in heat flux temperature dependence at radiator transition to the SC state.

In conclusion, we observe a strong suppression of the NF heat transfer between plane-parallel NbN layers of thickness $L$=270 nm due the superconducting transition of the absorber. This key result can be interpreted physically in terms of the TE mode (electrical field parallel to the surfaces), which is dominating the NF heat exchange between the plane-parallel surfaces of metals in the normal state. As distinct from the transverse magnetic (TM) mode, *the TE mode is strongly suppressed in the SC state* (possibly surrendering to the TM mode contribution), causing the pronounced decrease in the NF heat transfer, even when the radiator temperature is far above $T_c$. The results on NbN provide a significantly better correspondence between the measurements and the theory compared to the earlier study using elemental Nb [1], where the effect was recently discovered.

**Acknowledgements**

We thank D. Munzar, L. Skrbek, and A. Srnka for valuable discussions and support, P. Hanzelka for advices on experimental improvements, and M. Zobač and M. Kasal for the design of an electronic system for capacitance measurements. We acknowledge support by the Czech Science Foundation (GA CR) under Grant No. GA14-07397S and by the Czech Ministry of Education, Youth and Sports (MEYS), Project No. LO1212.

# Supplemental Material:
# Strong suppression of near-field radiative heat transfer by superconductivity in NbN

Authors: Věra Musilová, Tomáš Králík, Tomáš Fořt, Michal Macek
*Institute of Scientific Instruments of the CAS, v. v. i., Královopolská 147, 612 64 Brno, Czech Republic*

## Contents:

## 1. Optical constants of niobium nitride (NbN)

Optical constants are calculated using the computer code published in [1] for evaluation of dynamic electrical conductivity of a weak-coupling BCS superconductor. This code, based on Mattis and Bardeen theory [2], takes into account an arbitrary purity of a superconductor. In the normal state, the conductivity coincides with the Drude model

$$\sigma = \sigma_{DC} / (1 - i\omega\tau), \quad \tau = \sigma_{DC} / (\varepsilon_0 \omega_p^2) \tag{1}$$

where we used published value [3] of the plasma frequency $\omega_p = 1.47 \times 10^{16}$ rad/s. The calculated optical constants correspond to our measured values of the direct current conductivity $\sigma_{DC}$ = 8.5x10$^5$ S, superconducting critical temperature $T_c$ = 15.25 K and to the NbN energy gap inferred from published experimental data [4,5], by applying the relation $2\Delta_0 \approx 3.9 k_B T_c$. In the frequency region important for the heat fluxes at temperatures below 50 K, the corresponding very short relaxation time $\tau = \sigma_{DC} / (\varepsilon_0 \omega_p^2) \approx 4.5 \times 10^{-16}\,\mathrm{s}$ of NbN does not influence the real part of dynamic conductivity (which is decisive for absorption) and results also in a very high impurity parameter [1] $y = \hbar / 2\Delta\tau > 280 \;\; (T > 5\mathrm{K})$. This makes the model only weakly sensitive to the exact value of relaxation time (or the plasma frequency) both in normal and superconducting states.

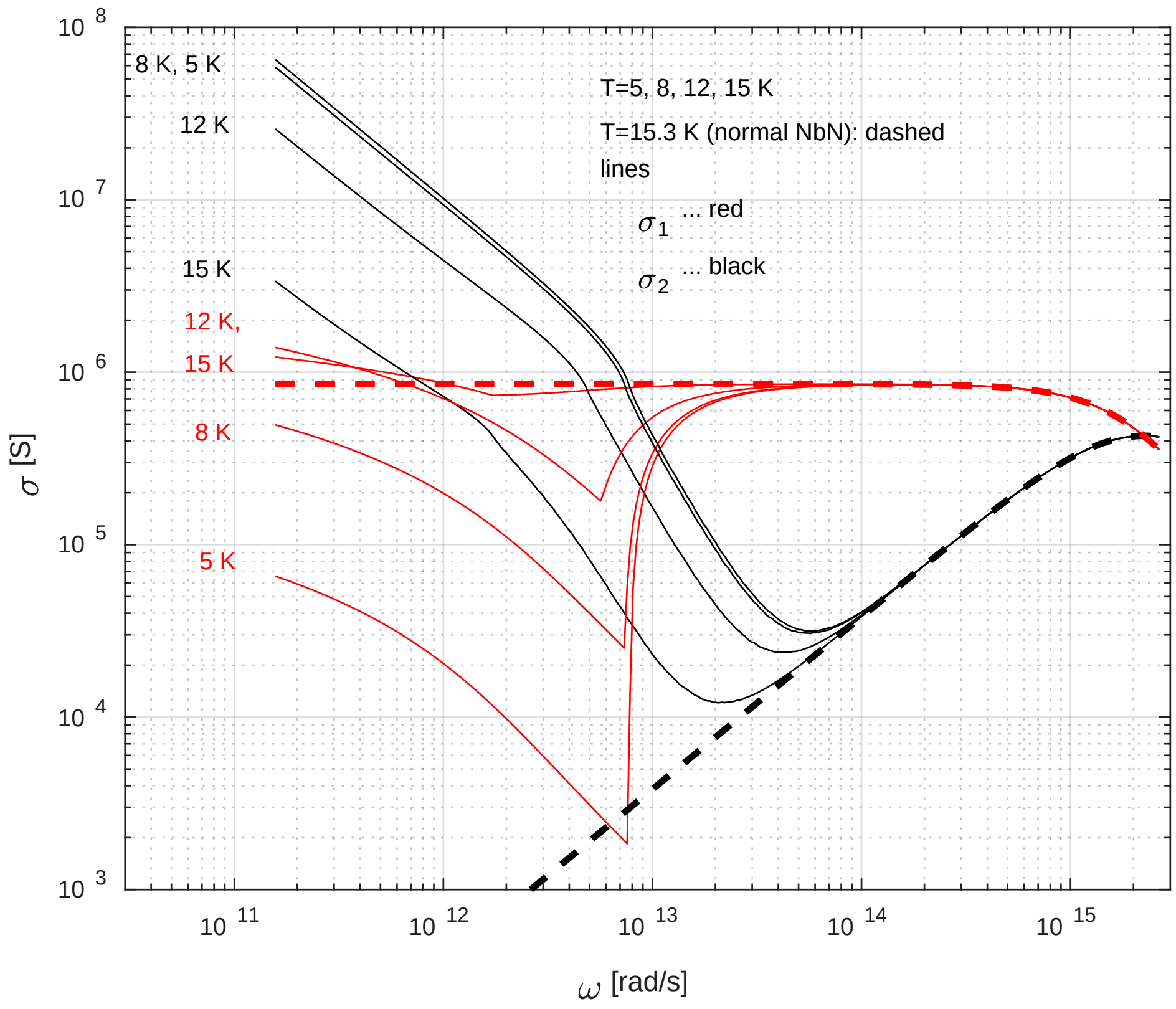

Fig. 1. Real ($\sigma_1$) and imaginary part ($\sigma_2$) of the dynamical conductivity of NbN. Dashed lines represent the Drude model with plasma frequency $\omega_p$ = 1.47x10$^{16}$ rad/s.

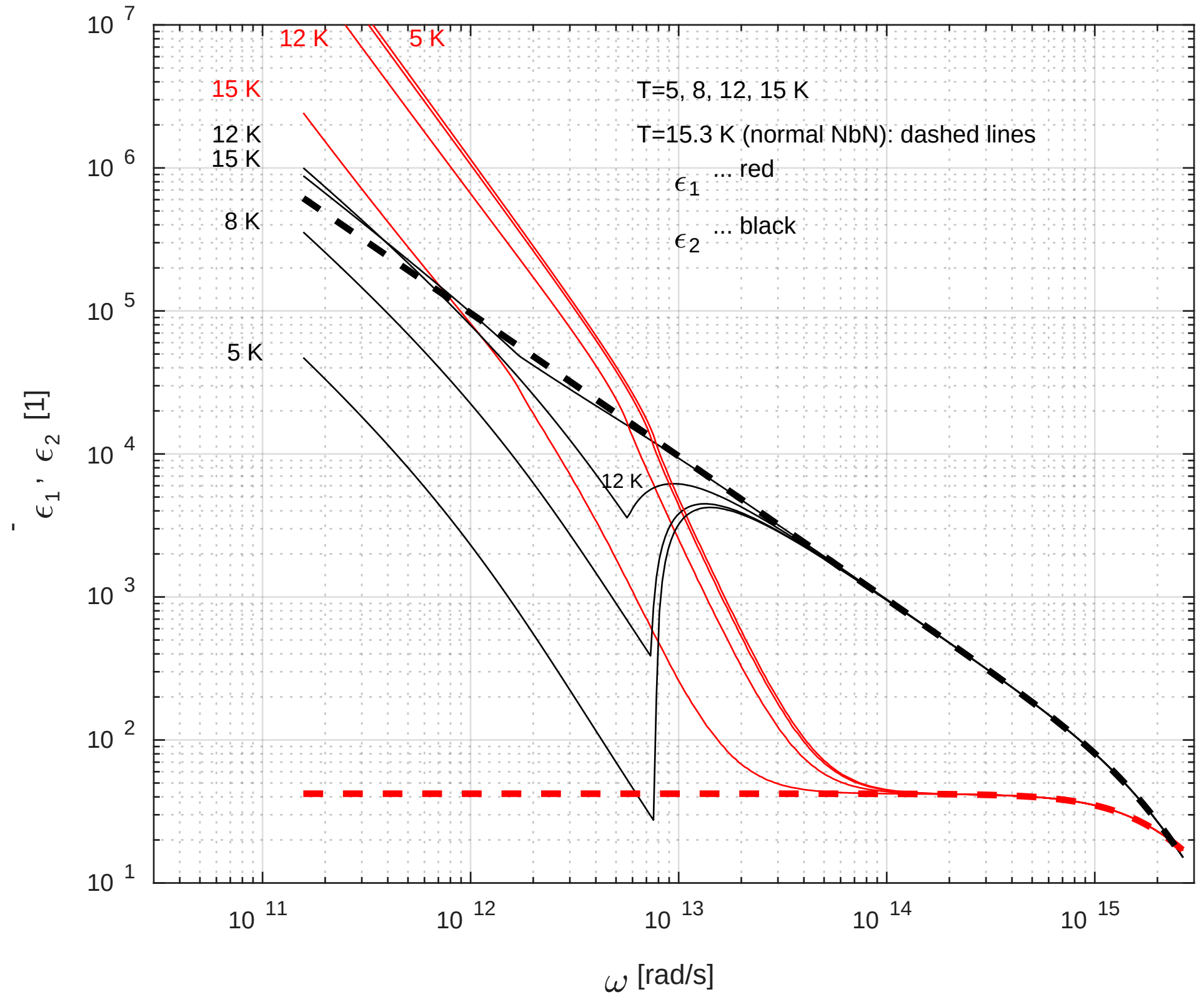

Fig. 2. Real ($\epsilon_1$) and imaginary part ($\epsilon_2$) of dynamical permittivity of NbN corresponding to conductivity values plotted in Fig. 1. Permittivity is calculated as $\varepsilon = 1 + \mathrm{i}\sigma/\varepsilon_0\omega$ .Dashed lines represent Drude model with plasma frequency $\omega_p$=14.7x10$^{15}$ rad/s.

## 2. Fresnel coefficients and "reflectivity" of NbN in far and near field

We calculate Fresnel coefficients of the NbN surface for both $K < K_0$ (far field) and $K > K_0$ (near field) as

$$r_{TE} = (\gamma_0 - \gamma)/(\gamma_0 + \gamma), \quad r_{TM} = (\varepsilon\gamma_0 - \gamma)/(\varepsilon\gamma_0 + \gamma), \quad \gamma_0 = \sqrt{K_0^2 - K^2}, \quad \gamma = \sqrt{\varepsilon K_0^2 - K^2} \tag{2}$$

where $K_0 = \omega / c$ is the wave vector in vacuum, $K$ is to the surface parallel component of the wave vector $K_0$ and $\varepsilon$ is the complex relative permittivity of the sample. Reflectivity of the surface reads

$$R = 0.5 * \left( |r_{TE}|^2 + |r_{TM}|^2 \right). \tag{3}$$

In Fig. 3a we plot the reflectivity $R$ in the $K_0$-$K$ space (corresponding essentially to the $\omega$-$K$ space via $K_0$=$\omega$/c) within the broad interval of frequencies covering the plasma frequency of NbN, where the "near field reflectivity" is dominated by the surface plasmon.

Nevertheless, for our low temperature experiment, T ≤ 50 K, the relevant frequencies are much lower, covering $K_0 < 10^5$ ($10^6$) rad/m, see Fig. 3b and 5. Within this region of frequencies, we also plot the real and imaginary parts of the Fresnel coefficients of NbN both in the normal state (Fig. 4a, 4b) and in the superconducting state at 5 K (Fig. 6a, 6b).

Contrary to the case of far field radiation, the thermal electromagnetic near-field is composed of evanescent waves which are created by thermal waves totally reflected within the sample at the sample-vacuum interface ($K > K_0$, $\mathrm{Re}(\gamma_0) = 0$). Important role in the near-field heat transfer is played by the imaginary part of Fresnel coefficients (see Eq. 4b in the paper or in sec. 3 of this Supplement).

Notice that in the normal state, the imaginary part of the TE-mode Fresnel coefficient dominates at low values of $K_0$ (low frequencies) in near field region, see Fig. 4b, left panel. In the superconducting state, this dominance is suppressed at lower $K$ values, while the values of Im($r_{TE}$) remain high at high $K$ (Fig. 6b). We remark that in the relation for the near field transmissivity of a vacuum gap, the higher values of imaginary parts of Fresnel coefficient at high $K$ are reduced by the multiplying exponential term (see Eq. 4b in Sect. 3 of this Supplement), $exp(-2d\sqrt{K^2 - K_0^2}) \approx exp(-2Kd)$.

## *Normal state of NbN*

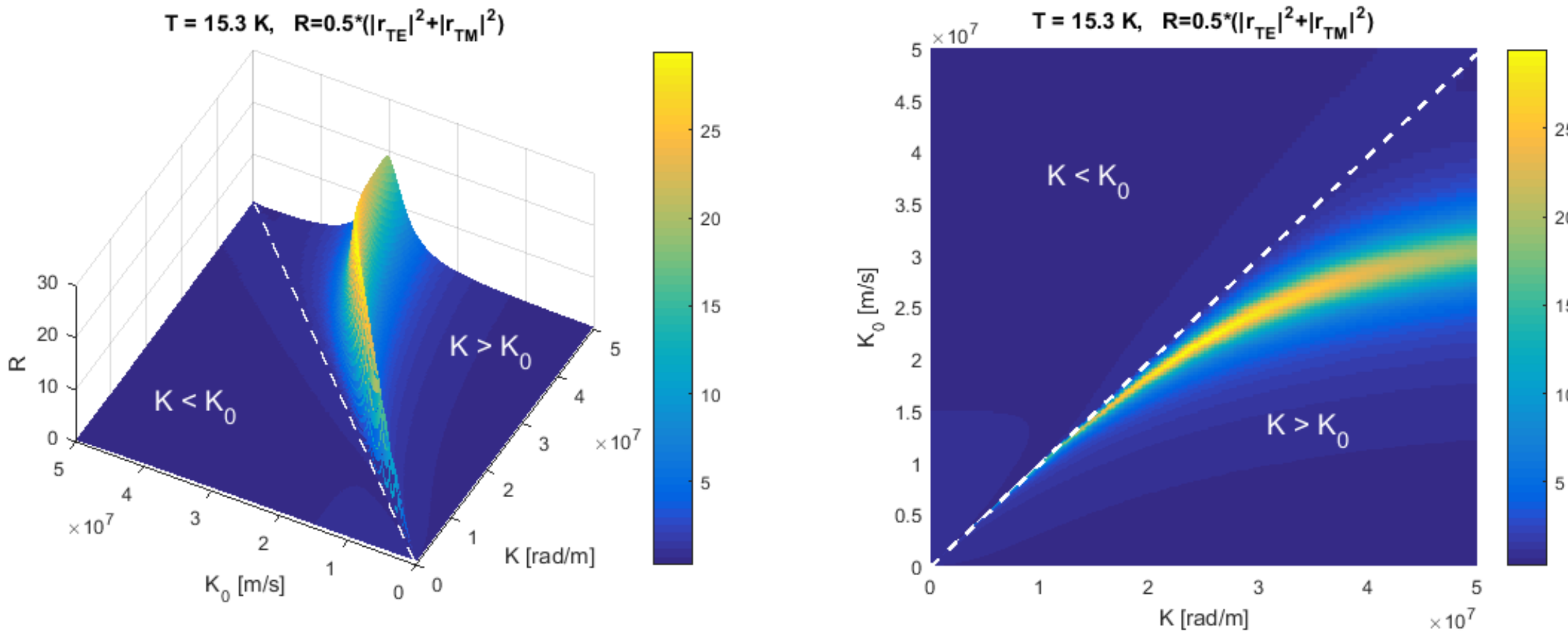

Fig. 3a. **Reflectivity** $R = (|r_{TE}|^2 + |r_{TM}|^2)/2$ calculated for a broad interval of frequencies $\omega = cK_0$ (c=3x10$^8$ m/s), up to bulk plasma frequency $\omega_p \approx 1.5\times10^{16}$ rad/s . Dashed lines separate far field ($K < K_0$) from the near field ($K > K_0$) region. Surface plasmon "approaching" to its resonance frequency at $\omega = \omega_p/\sqrt{2}$ ($K_0 \approx 3\times10^7$ rad/m) is visible. For excitation of this resonance a temperature of thousands Kelvins would be necessary.
*Left panel*: 3D view of $R$; *right panel*: 2D view of $R$.

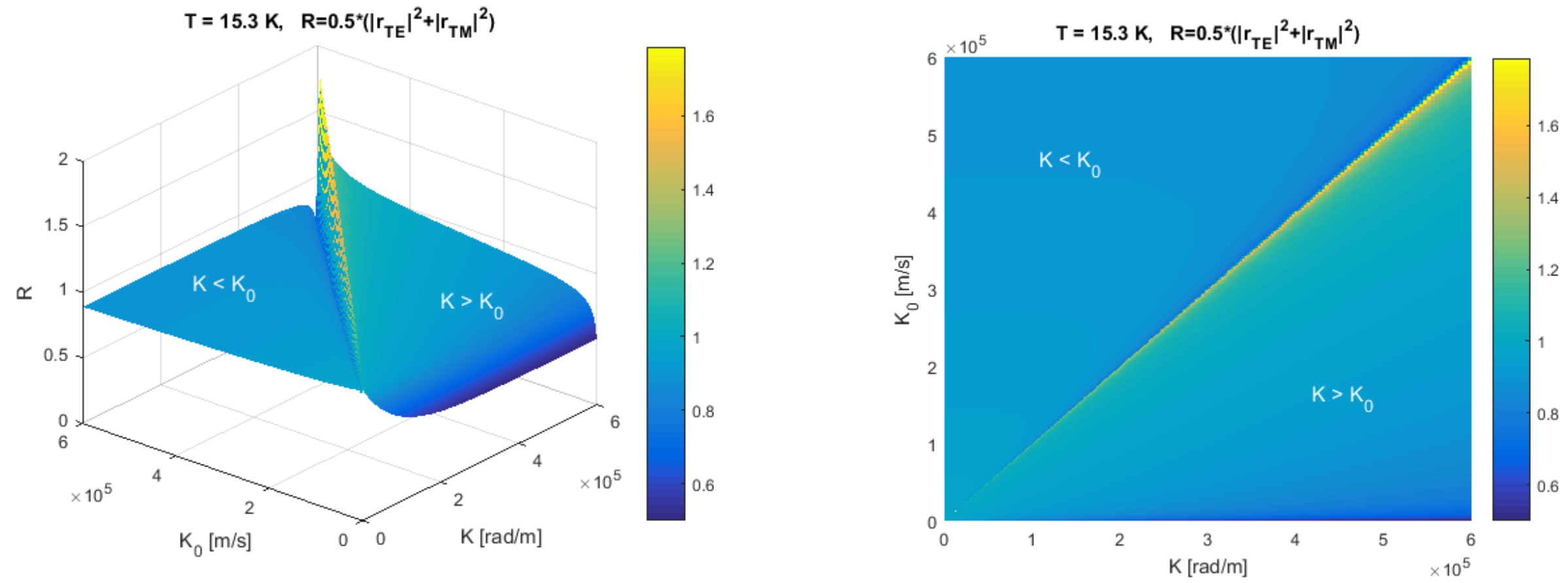

Fig. 3b. **Reflectivity.** The same as in Fig. 3a, but for lower frequencies, which are relevant to experiment at temperatures T < 50 K ($K_0 < 6\text{x}10^5$ rad/m, and wave vectors $K < 6\text{x}10^5$ rad/m).

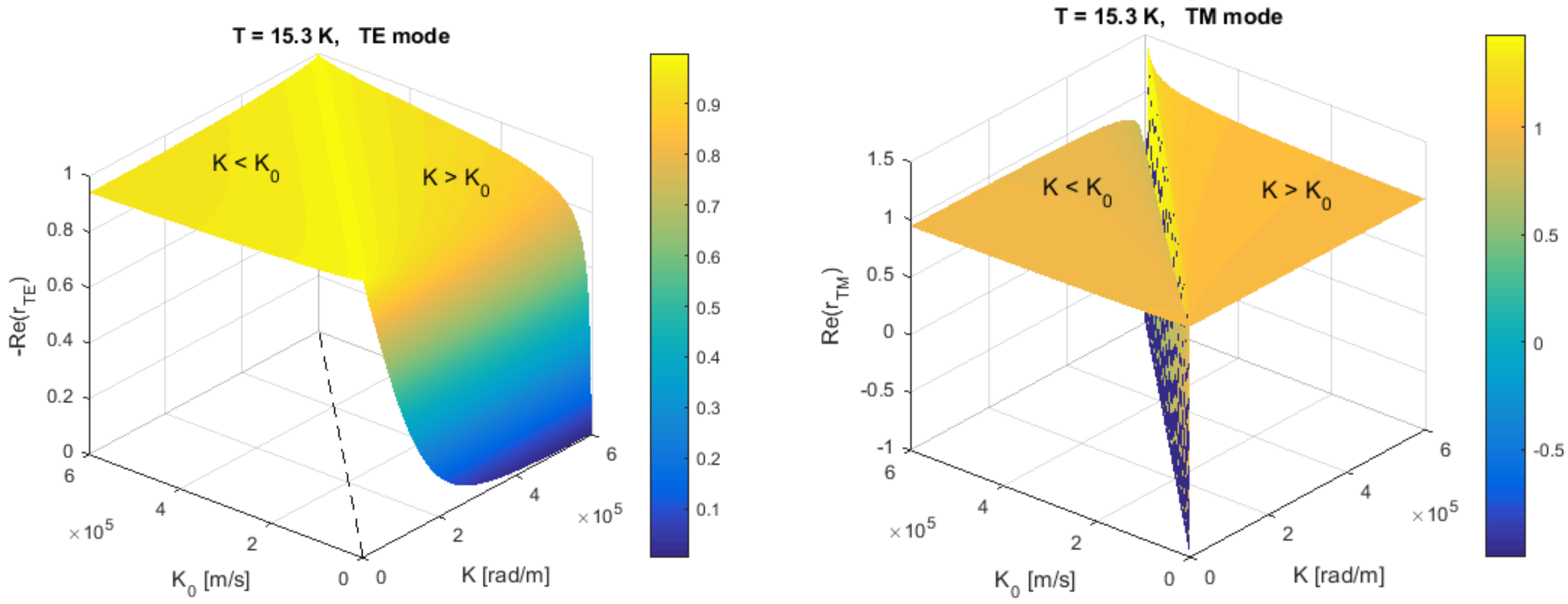

Fig. 4a. **Real part of the Fresnel coefficients**. TE mode (left), TM mode (right).

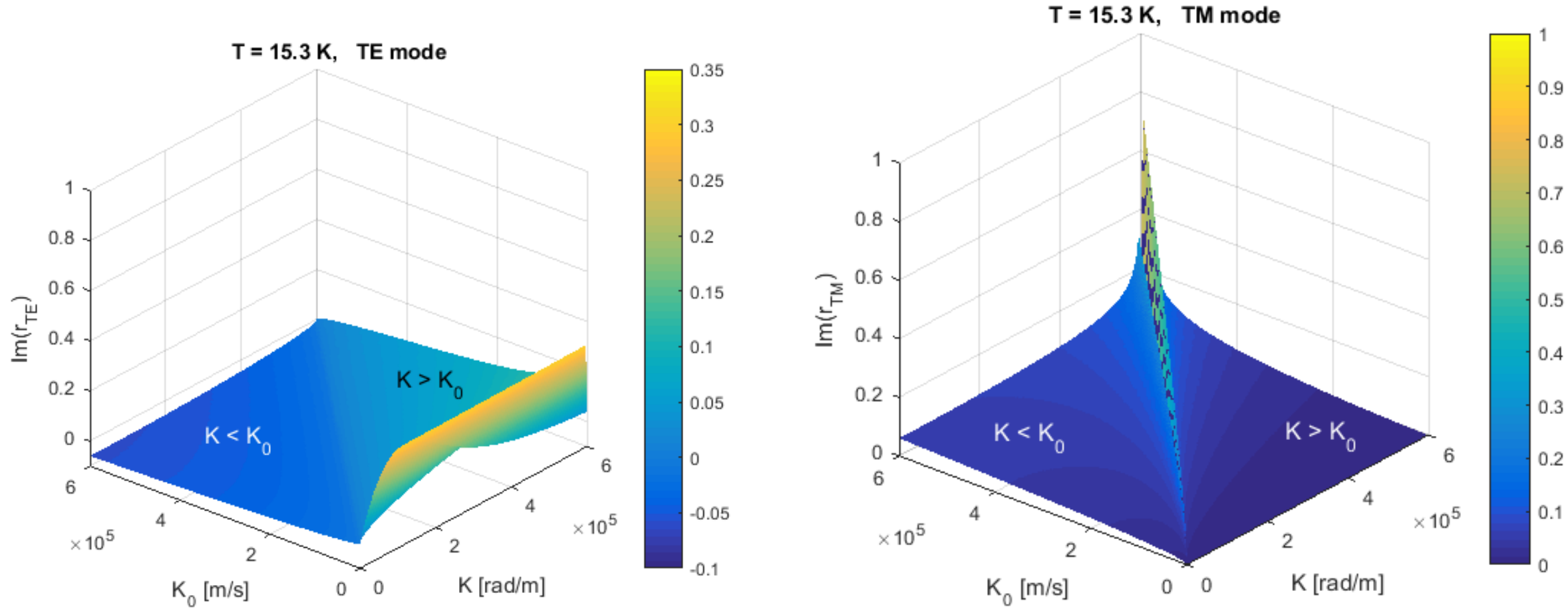

Fig. 4b. **Imaginary part of the Fresnel coefficients**. TE mode (left), TM mode (right). Notice the low frequency maximum in Im($r_{TE}$) along broad span of wave vectors $K > K_0$.

## *Superconducting state of NbN*

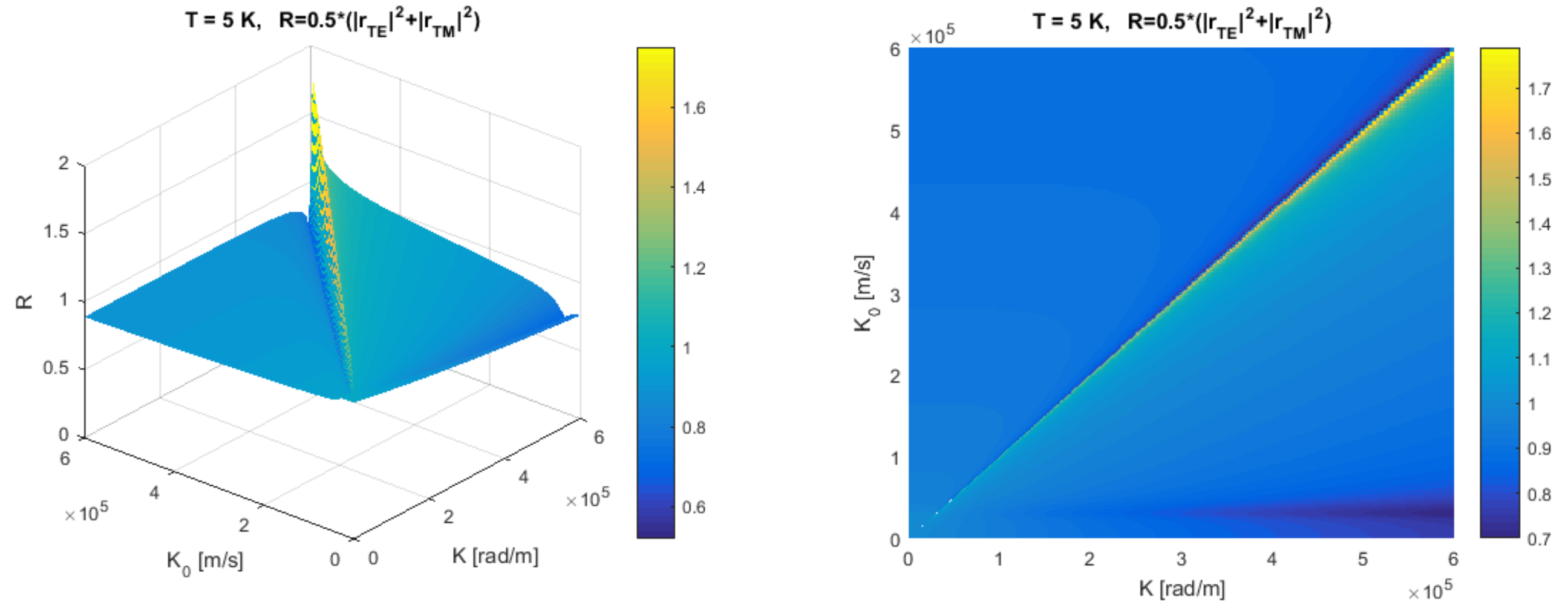

Fig.5. **Reflectivity.** For frequencies which are relevant to experiment at temperatures T < 50 K ($K_0 < 6x10^5$ rad/m, and wave vectors K < $6x10^5$ rad/m). The superconducting gap at zero temperature is at $K_0 = 2.6x10^4$ rad/m.

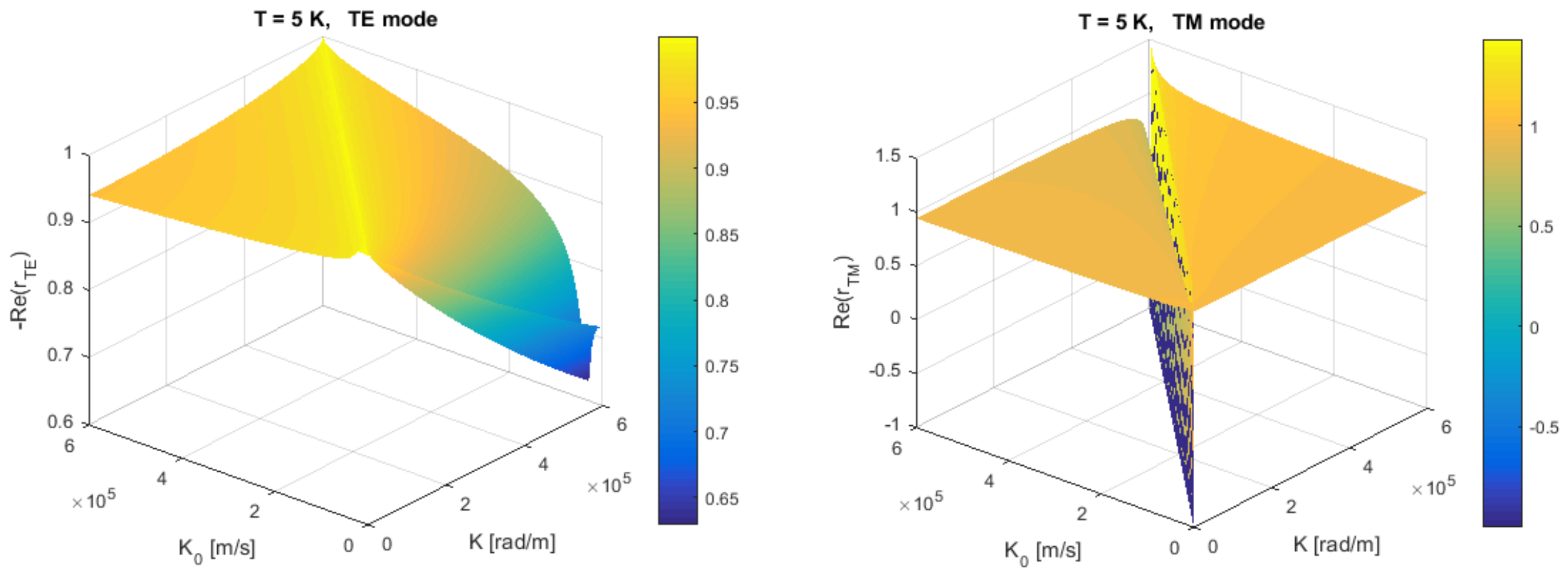

Fig. 6a. Real part of the Fresnel coefficients. TE mode (left), TM mode (right).

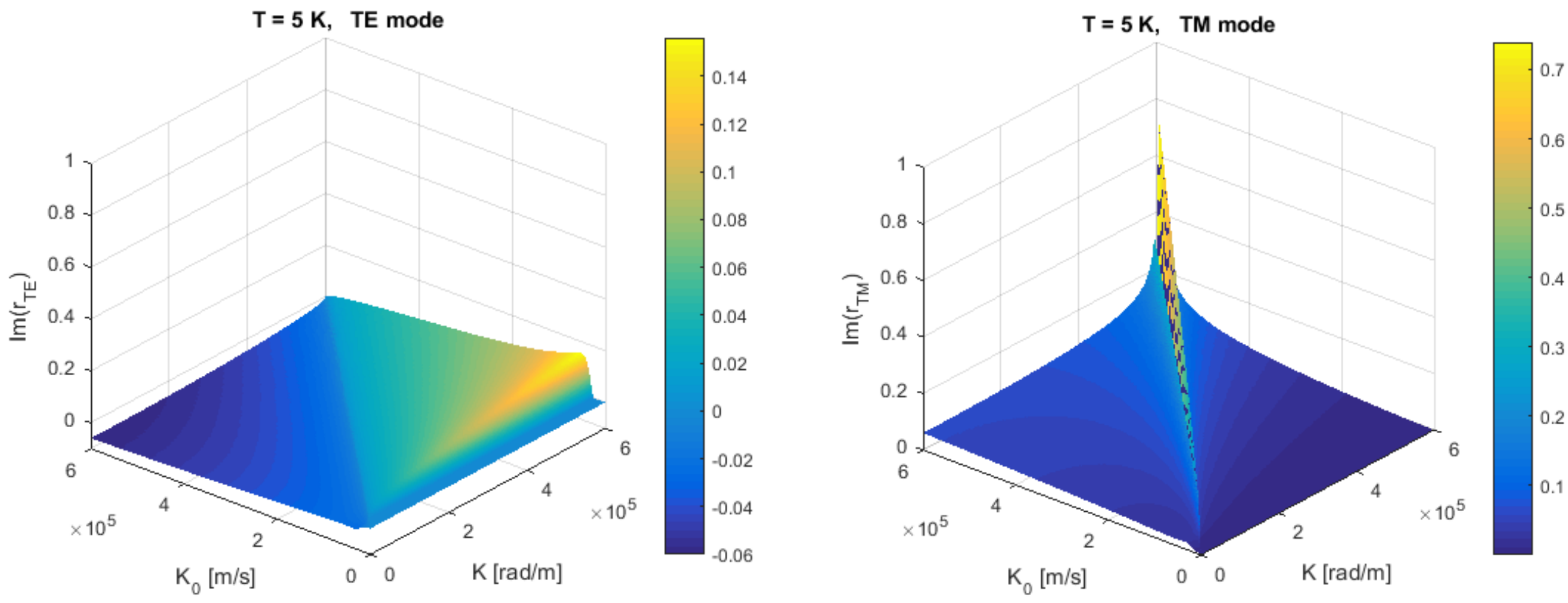

Fig. 6b. **Imaginary** parts of the Fresnel coefficients. TE mode (left), TM mode (right). Notice the suppression of Im($r_{TE}$) at low frequencies $K_0$ (compare with Fig. 4b). The superconducting gap at zero temperature is at $K_0 = 2.6x10^4$ rad/m.

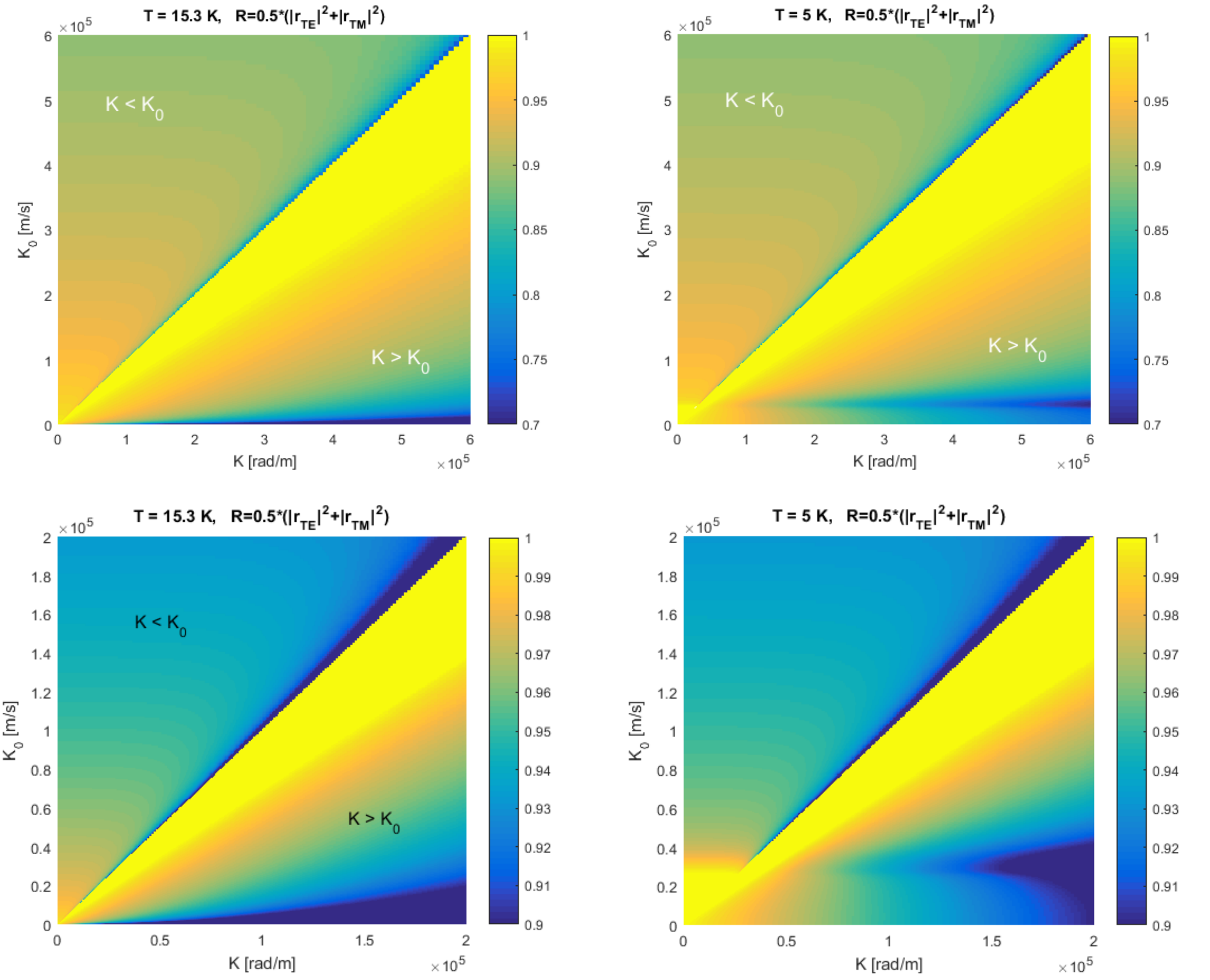

Fig.7. Comparison between reflectivity $R$ in the normal (left) and superconducting (right) states. Note that the ($K$, $K_0$) space is zoomed below 6x10$^5$ rad/m (upper panels) and below 2x10$^5$ rad/m (lower panels). At $K < K_0$ (far field region, where $R < 1$), we see a little increase (by 1-3%) of reflectivity within superconducting (SC) gap ($K_0 < 2.6\text{x}10^4$ rad/m). Within the SC gap at $K > K_0$, the reflectivity increases above $R = 0.9$ up to $K$=2x10$^5$ rad/m and above $R = 0.7$ up to long wave vectors $K$=6x10$^5$ rad/m. *R values lower than 0.7 (upper panels) or 0.9 (lower panels) are visualized by dark blue. The values of* $R \geq 1$ *in the region* $K > K_0$ *are yellow.*

## 3. Transmissivity of vacuum gap between NbN surfaces

Examples of transmissivity $t(K, K_0)$ of the vacuum gap between plane parallel surfaces were calculated from Eqs. (4a) and (4b), where $K_0 \equiv \omega/c$ is a wave vector in vacuum and $K$ is its projection to the planes limiting the gap.

$$t_m^{FF} = \frac{\left(1-\left|r_m^{(1)}\right|^2\right)\left(1-\left|r_m^{(2)}\right|^2\right)}{\left|1-r_m^{(1)}\, r_m^{(2)}\exp(2\,\mathrm{i}\gamma_0 d)\right|^2} \qquad \text{(4a, 4b)}$$

$$t_m^{NF} = \frac{4\,\mathrm{Im}(r_m^{(1)})\,\mathrm{Im}(r_m^{(2)})\exp(-2\gamma_0'' d)}{\left|1-r_m^{(1)} r_m^{(2)}\exp(-2\gamma_0'' d)\right|^2}, \quad m = TE, TM$$

We define the "reduced transmissivity X" derived from the relation (4b) for near field transmissivity, in which the exponential factor $\exp(-2\gamma_0'' d)$ in the numerator is omitted:

$$X_m \equiv \frac{4\,\mathrm{Im}(r_m^{(1)})\,\mathrm{Im}(r_m^{(2)})}{\left|1 - r_m^{(1)} r_m^{(2)} \exp(-2\gamma_0'' d)\right|^2}, \quad m = TE, TM \tag{5}$$

This quantity is plotted in Fig. 8 for both the TE and TM modes and for normal as well as the SC state of the absorber. Again as in Fig. 3a, we can see the surface plasmon resonance in $X_{TM}$ in the TM mode at $K_0 \approx 3\times10^7$ rad/m ($\omega \approx \omega_p/\sqrt{2}$). Nevertheless, this resonance does not contribute to the heat transfer in low temperature experiments.

Transmissivities of the vacuum gap between identical plane parallel NbN surfaces are calculated for three different models of the sample. Refraction index of sapphire substrate is approximated by index of the ordinary ray in each case:

1) Sample model A ("Real samples"): the sample is modeled by a 270nm thick NbN layer on a 2.7 mm thick plane-parallel sapphire substrate with semi-infinite metal attached to the reverse side of the sapphire, simulating metallization of that side of substrate.
2) Sample model B (NbN on semi-infinite sapphire substrate)
3) Sample model C (NbN in vacuum): both NbN layers are placed in vacuum without any substrate. This model together with A and B is tested in section 4.

## *Reduced transmissivity X of vacuum gap with model samples B*

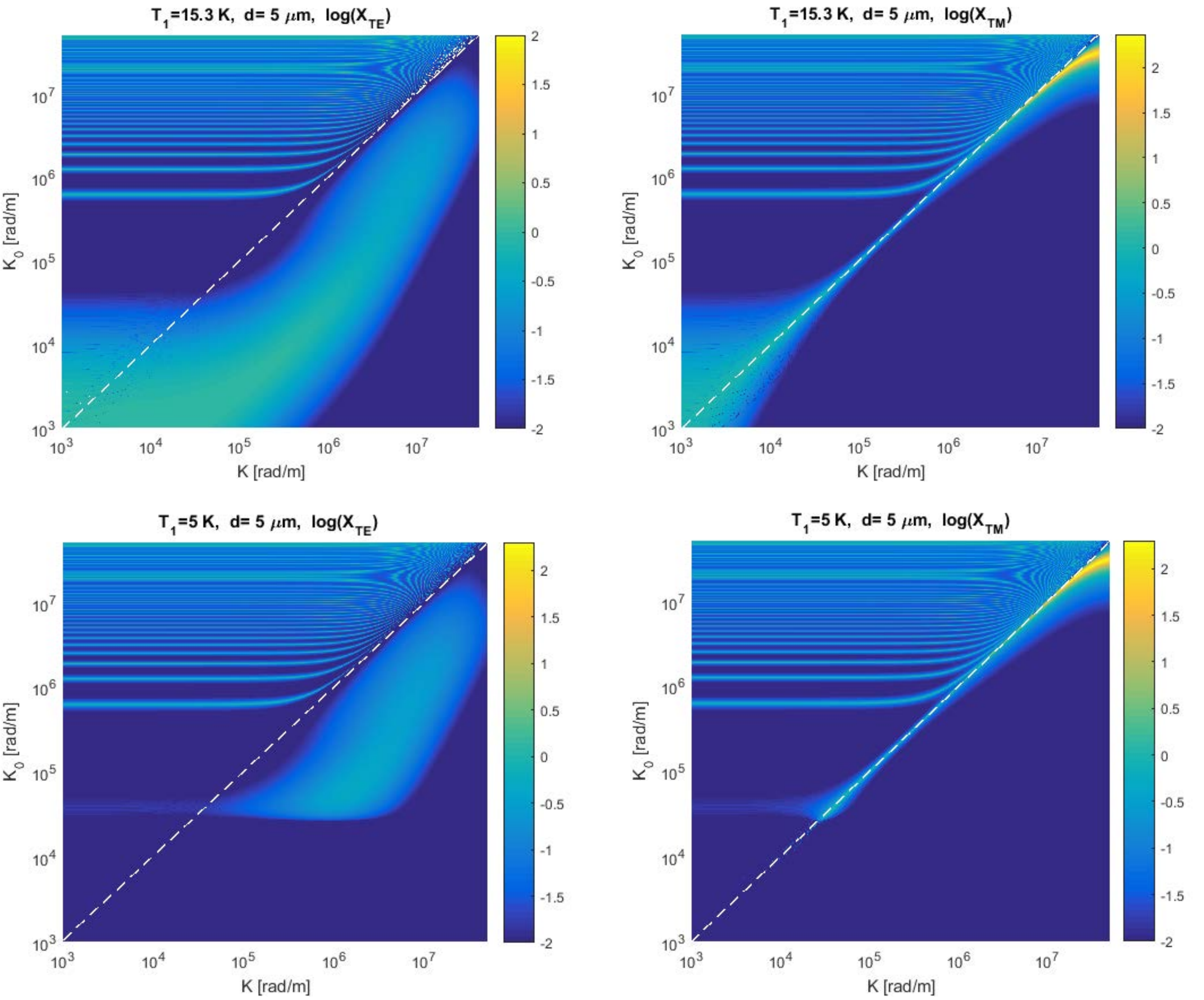

Fig. 8. **Reduced transmissivity *X* of the vacuum gap** calculated from Eq.(5) **with sample model B**. In the far field region ($K < K_0$), the interference bands are mutually displaced by $\Delta K_0 = 10$ µm = 2*d.* In the TM mode near field transmissivity, we see the surface plasmon resonance at $K_0 \approx 3\text{x}10^7$ rad/m.

## *Transmissivity of the vacuum gap between “real samples” (sample model A)*

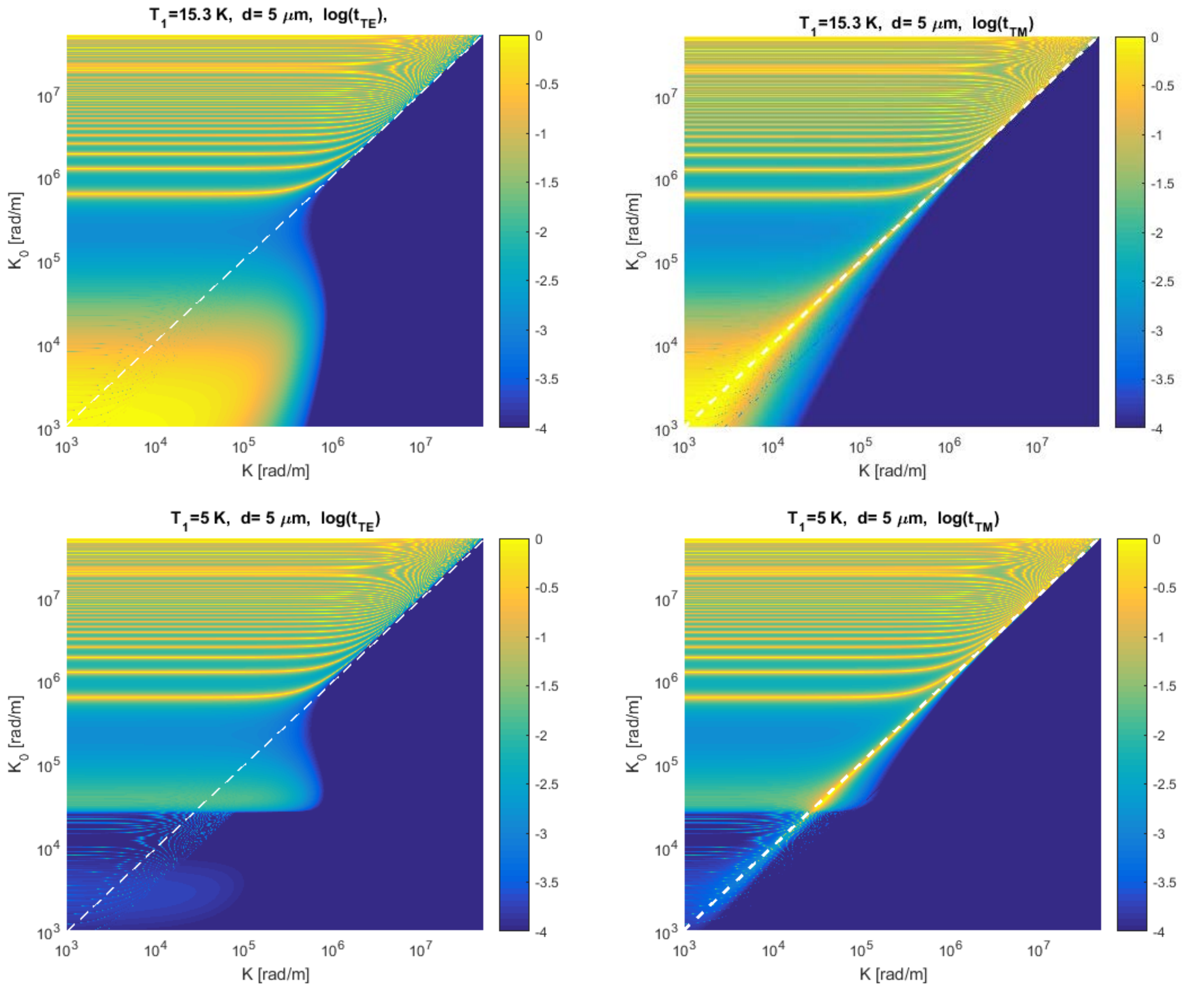

Fig. 9a. **Transmissivity of vacuum gap with sample model A.** Bright yellow bands are due to interference of far field radiation in the vacuum gap, visible both in normal and superconducting state. Thin light blue interference bands reveal presence of interference within sapphire substrates, recognizable also at low values of $K_0$ and $K$ in near field region of vacuum gap. Notice the effect of superconducting gap at $K_0 < 2.6 \times 10^4$ rad/m

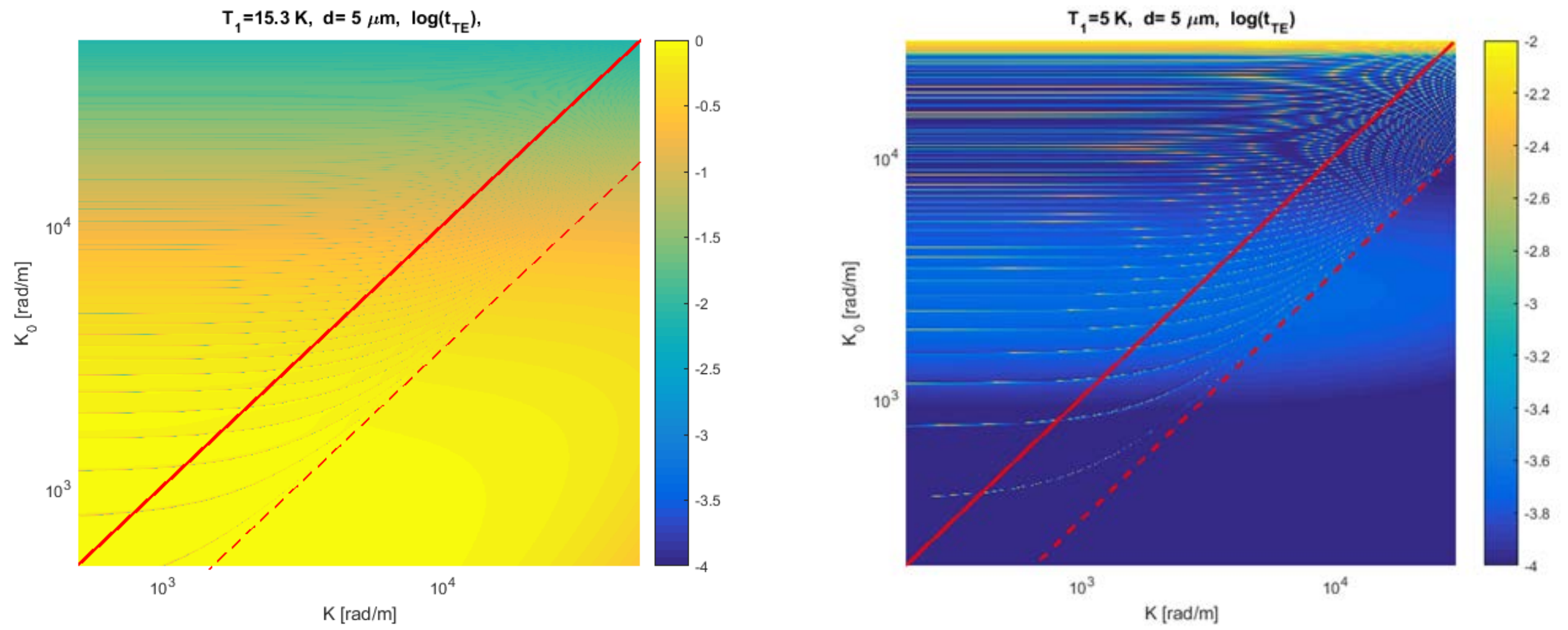

Fig. 9b. **Detail of left panels in Fig. 9a.** Thin interference bands recognizable at low $K$ values are distant by $\Delta K_0$=380 rad/m ($\Delta\lambda_0=2L$, where $L$=2.7mm is the sapphire substrate thickness; index of refraction $\mathrm{Re}(N_S) \approx 3.07$). The bands overlap part of the vacuum gap near field region ($K_0 < K$), being limited by the condition $K = \mathrm{Re}(N_S)K_0$, red dashed line.

### *Transmissivity of the vacuum gap with sample model B*

This model excludes interference on the sapphire substrate which is now semi-infinite. Comparing transferred heat fluxes in the sample model A (see transmissivities in Fig. 9a, lower panel) and B (Fig. 10a), we see that the variation in sapphire substrate has little effect on transferred heat flux. For hot sample at 20 K, cold sample at 5 K and vacuum gap 5 μm, the dominating near field heat fluxes in A and B model differ by less than 0.5%.

| | | |
|---|---|---|
| "Real sample" (Sample model A): | $q_{TE}$= 0.0019956 W/m2, | $q_{TM}$= 0.00089985 W/m2 |
| Semi-infinite substrate (Sample model B): | $q_{TE}$= 0.0019963 W/m2, | $q_{TM}$= 0.00090409 W/m2 |

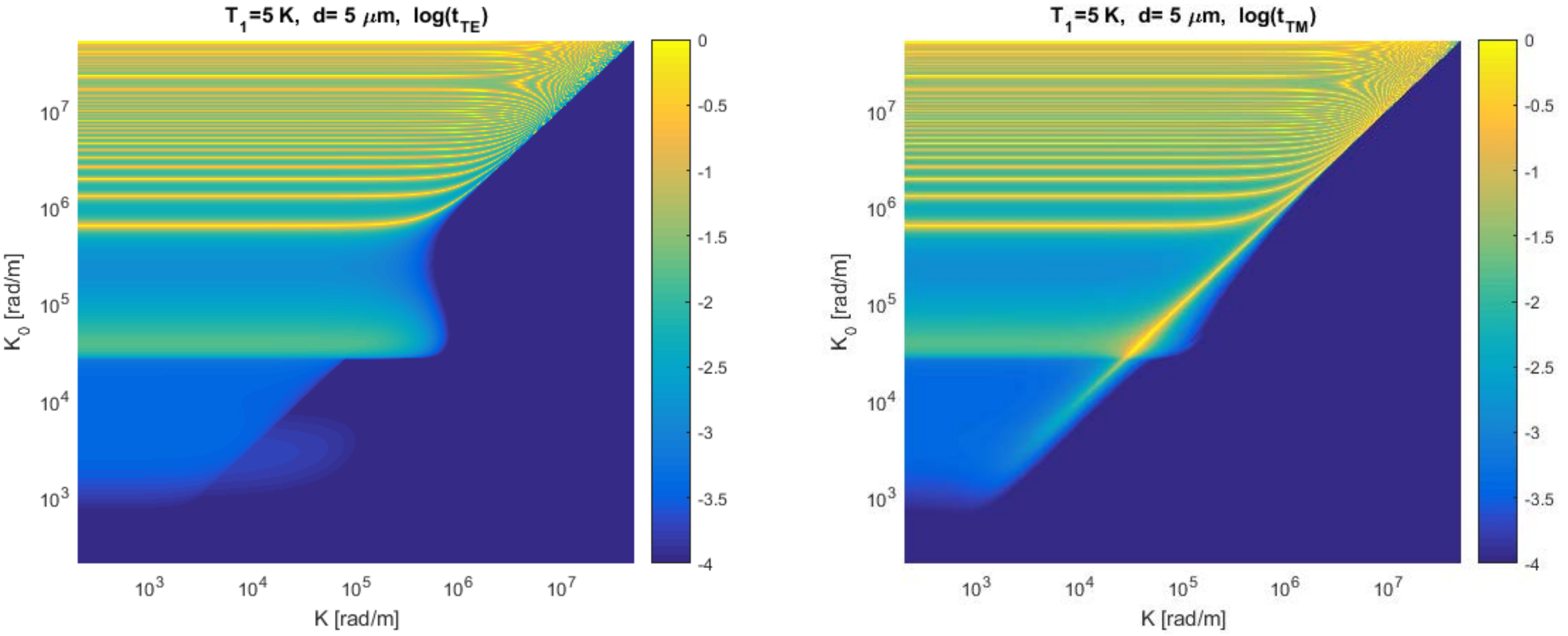

Fig. 10a. **Transmissivity of the vacuum gap between samples B**. Superconducting state of colder NbN layer.

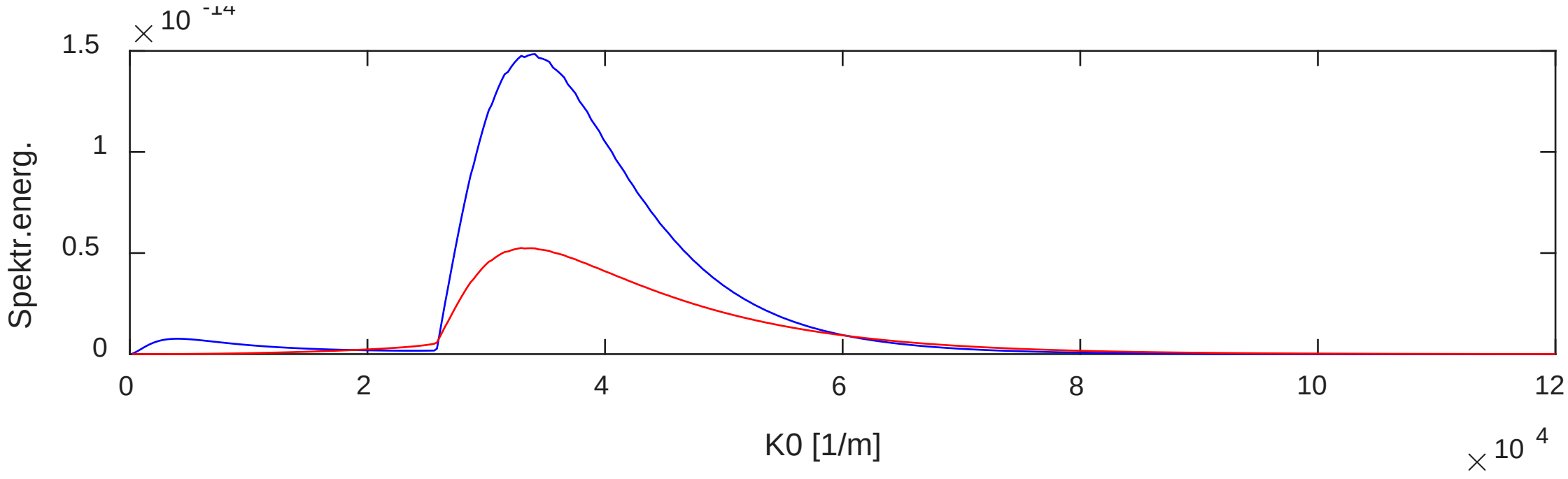

Fig. 10b. **Frequency spectrum of the near field heat flux** derived from the transmissivities in Fig. 10a by integration over the wave vector *K* states. Planck's black body law corresponding to T = 20 K was used. TE mode (blue), TM mode (red) are shown separately.

## 4. Sensitivity of the sample model to parameter variation

To evaluate the sensitivity of heat flux to the measured parameters of the NbN samples, we performed several dedicated calculations of heat flux $q$ transferred over vacuum gap aimed to verify correctness of the theoretical model. The parameters in question are the thickness $L$, the SC energy gap [i.e. $n$ and $T_c$ in Eq. (10)], the electrical conductivity $\sigma_{DC}$ and NbN plasma frequency value (or electron relaxation time). We evaluate also the effect of the substrate.

Tests were performed at two representative values of the vacuum gap spacing and two absorber temperatures: $d$=5 μm and 15 μm, $T_1$=5 K (SC) and 15 K (normal metal). The radiator temperature was fixed at $T_2$=20 K, where the maximum contrast was observed (see Fig. 11c).

## *Thickness of NbN layer and the substrate effect*

In panels (a) and (b), the Fig. 11 plots the dependence of heat flux on the NbN layer thickness *L* for two samples: (i) an NbN layer on top of a 2.7 mm thick sapphire substrate, covered with a reflective metallic coating on the reverse side of the sapphire substrate and (ii) an identical NbN layer, but without any substrate. We note that in experiments for very thin layers of NbN the electrical conductivity and $T_c$ tend to decrease with thickness strongly, thus the results in Fig. 11 should be taken as rather academic for the thicknesses less than a few tens of nanometers. Importantly, the substrate effect is negligible for *L*>100 nm: There are little differences among the computed spectral (not shown) and total heat fluxes (Fig. 11 a, b) whether we modelled the realistic samples sputtered on the sapphire substrate (Sample model A), or fictitious NbN layers placed in vacuum (Sample model C).

Fig. 11, panel (c), plots the contrast between heat flux with the absorber in normal and SC states, derived from the dependences shown in Fig. 11 (a) and (b). For the vacuum gaps *d*=5 µm and *d*=15 µm wide, the maximum theoretical contrast is achieved with the sample thicknesses of *L*≈250 nm and *L*≈500 nm, respectively.

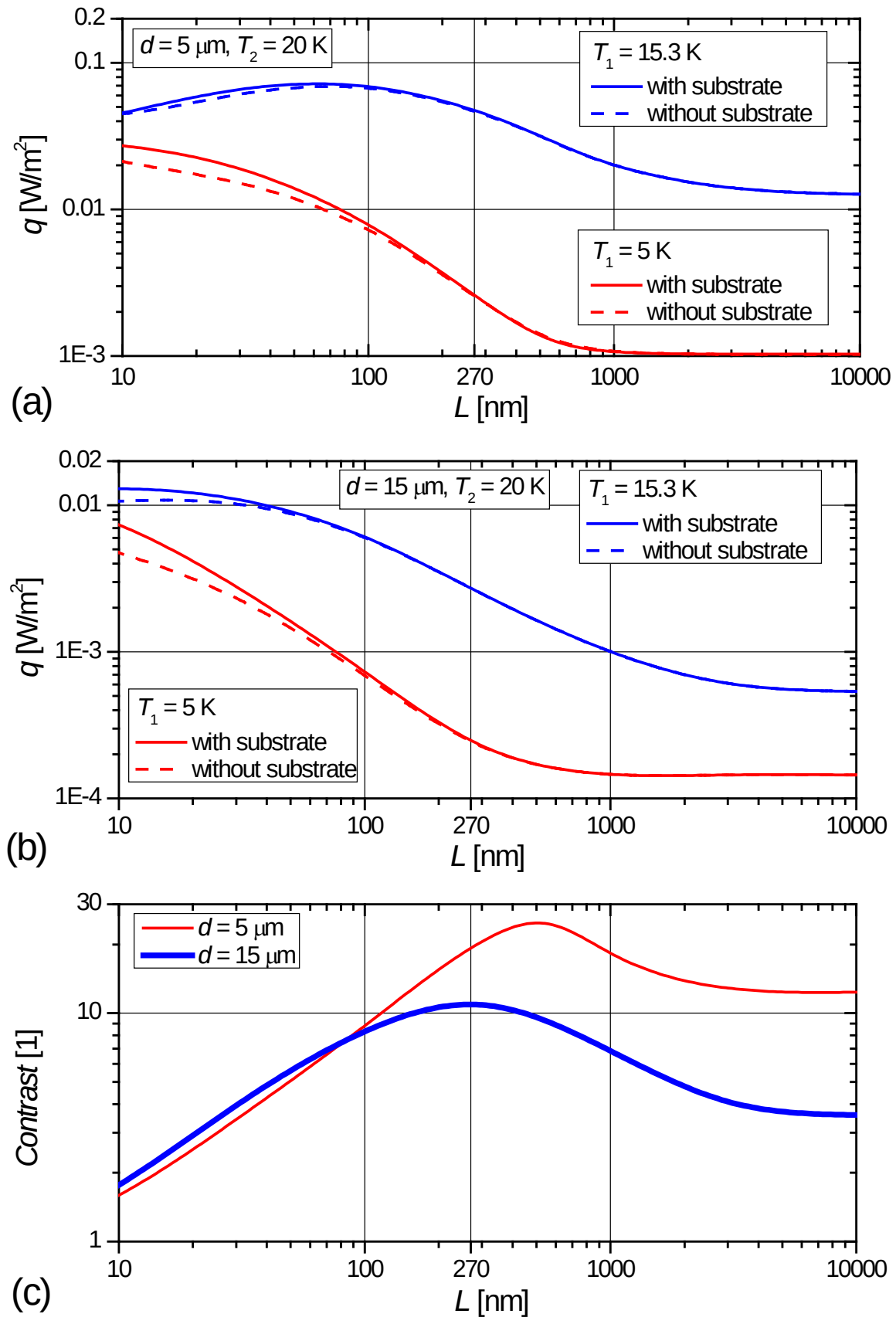

**Fig. 11. The dependence of heat flux *q*** (panels a, b) and contrast *C*, Eq. 14, (panel c) on NbN layer thickness *L* at two vacuum gaps. The *q* dependence for absorber in normal (upper solid line) and SC state (lower solid line) are plotted for vacuum gap d=5 µm (a) and d=15 µm (b). Full lines are for model of real sample (sapphire substrate metallized from reverse side), while dashed lines are for NbN layer in vacuum without any substrate. The *L*-dependences of the contrast *C* (panel c) show maxima near the samples thickness 270 nm used in experiment.

## *Energy of the SC gap and the DC conductivity of NbN*

To evaluate sensitivity of heat flux to the value of the SC energy gap $E_{g0}$, we used the relation $\hbar\omega_{g0} = 2nkT_c$ with the BCS value *n*=1.764 and compared it with results presented in the main article [*n*=1.95, Eq. (10)] in the calculation of *d*-dependences at $T_2$=20 K and $T_1$=5 K. We obtained the following approximate variation of *q*(*d*) with *n*:

$$\delta q / q \approx -3\delta n / n \approx -3\delta T_c / T_c \qquad (6)$$

where the $T_c$-sensitivity is the same as *n*-sensitivity by virtue of the relation $E_{g0}=2nkT_c$.

The main question is whether we can correct the theoretical heat flux dependence *q*(*d*), surpassing the experimental one by a factor of 1.5-2 with the absorber in the SC state (see Sec. V of the main article). To do that, we would need to decrease *n* or $T_c$ to improbable values: The *n* value should be decreased below the BCS value 1.764, which contradicts the experimental *n* values obtained in other laboratories in optical measurements (typically n≈2 or higher). The needed decrease in critical temperature, down to 13 K or below, to achieve the same effect is also very high.

For the same dependence of *q*(*d*), we compared the results for $\sigma'_{DC}=\sigma_{DC}/2$ and $\sigma_{DC}$ of the NbN electrical conductivity for both normal and SC state of absorber and came to the common approximate conclusion

$$q'/q \approx \sigma_{DC}/\sigma'_{DC} = \rho'_{DC}/\rho_{DC} . \qquad (7)$$

To increase the values in the heat flux dependence *q*(*d*) by a factor 1.5-2 would mean to increase resistivity of the sample by the same factor. Moreover, the common increase of heat flux in both SC and normal states contradicts to the fact that systematic excessive heat flux was not observed in normal state.

## *Plasma frequency of NbN (electron relaxation time)*

Measured dc conductivity $\sigma_{DC} = 8.5\times10^5\,\mathrm{S}$ and plasma frequency $\omega_p = 1.47\times10^{16}\,\mathrm{rad/s}$ taken from literature result in very short electron relaxation time $\tau = \sigma_{DC}/(\varepsilon_0\omega_p^2) \approx 4.5\times10^{-16}\,\mathrm{s}$ giving a high value of impurity parameter $y = \hbar/2\Delta\tau > 280\ (T > 5\mathrm{K})$ for the studied NbN samples. This makes the model only weakly sensitive to the exact value of relaxation time (or the plasma frequency). For example, comparing heat fluxes calculated for this relaxation time (y=280) and for a ten times longer value (y=28) calculated for $T_2$=20 K and $T_1$=5 K, we obtain still less than 5 % differences at all vacuum gaps used in the experiment.

## *Conclusion of section 4*

From the presented analysis, we can see that to explain the excessive heat flux observed with the superconducting absorber, we would need unrealistic variations of parameters in the sample model: Either to use NbN layer thickness *L* < 200 nm (instead of 270 nm), or the parameter *n* less than the BCS value *n* = 1.764 (instead of *n* = 1.95 used) or the critical temperature $T_c$ < 13 K (instead of 15.2 K actually measured) or the DC conductivity to be twice lower than the one actually measured. Model of the experiment is also little sensitive to the exact value of electron relaxation time (to plasma frequency). We thus cannot explain the measured excessive heat flux (differing by a factor of 1.5-2 from the theory) simply by uncertainties in the model parameters.